\documentclass[%
 aip,
 amsmath,amssymb,
 reprint,%
]{revtex4-1}

\usepackage{amsmath} %This was added by the authors
\usepackage{multirow} %This was added by the authors
\usepackage{xcolor} %This was added by the authors

\usepackage{graphicx}% Include figure files
\usepackage{dcolumn}% Align table columns on decimal point
\usepackage{bm}% bold math
\usepackage[utf8]{inputenc}
\usepackage[T1]{fontenc}
\usepackage{mathptmx}
\usepackage{etoolbox}

\makeatletter
\def\@email#1#2{%
 \endgroup
 \patchcmd{\titleblock@produce}
  {\frontmatter@RRAPformat}
  {\frontmatter@RRAPformat{\produce@RRAP{*#1\href{mailto:#2}{#2}}}\frontmatter@RRAPformat}
  {}{}
}%
\makeatother
\begin{document}

\preprint{AIP/123-QED}

\title[Full-field rheo-optical analysis of wormlike and networked micellar structures under uniaxial extensional flow]{Full-field rheo-optical analysis of wormlike and networked micellar structures under uniaxial extensional flow}
% Force line breaks with \\

\author{Masakazu Muto}
%\altaffiliation{
 \affiliation{ 
 Department of Mechanical Engineering, Nagoya Institute of Technology, Nagoya, Japan. Gokiso Showa-ku, Nagoya-shi, Aichi 466-8555, Japan. 
 }%Lines break automatically or can be forced by \\
 
\author{Naoki Kako}%
\affiliation{ 
Department of Mechanical Engineering, Nagoya Institute of Technology, Nagoya, Japan. Gokiso Showa-ku, Nagoya-shi, Aichi 466-8555, Japan.
}%
\author{Tatsuya Yoshino}%
\affiliation{ 
Department of Mechanical Engineering, Nagoya Institute of Technology, Nagoya, Japan. Gokiso Showa-ku, Nagoya-shi, Aichi 466-8555, Japan.
}%
\author{Shinji Tamano}
\affiliation{ 
Department of Mechanical Engineering, Nagoya Institute of Technology, Nagoya, Japan. Gokiso Showa-ku, Nagoya-shi, Aichi 466-8555, Japan.
}%

\date{\today}% It is always \today, today,

\begin{abstract}
The present study proposes a novel “full-field extensional rheo-optical technique” to investigate the relationship between the rheological properties and internal structural deformation of complex fluids under uniaxial extensional flow.
Macroscopic viscoelasticity from rheological measurements and microscopic birefringence from optical measurements are integrated to evaluate the microstructural deformation and orientation of the fluids under extensional stress.
The proposed technique integrates a liquid dripping method with a high-speed polarization camera to measure the extensional stress and flow birefringence simultaneously.
In the liquid dripping method, temporal evolution images of the liquid filament diameter for fluids dripping from a nozzle are measured to obtain the extensional stress loading on the filament.
These images are acquired using the high-speed polarization camera that captures full-field two-dimensional (2D) birefringence with high spatiotemporal resolution.
Wormlike and networked micellar solutions of cetyltrimethylammonium bromide (CTAB) and sodium salicylate (NaSal) with varying concentrations of CTAB and NaSal are employed as the measurement targets. 
Consequently, we successfully visualized temporally developing images of the flow birefringence field of uniaxially extending micellar solutions induced by the orientation of micelles.
Furthermore, the proposed technique supports investigating the conditions for establishing the stress-optical rule, which is the linear relationship between stress and birefringence.
The stress-optical coefficient, which is a proportionality constant indicating the sensitivity of birefringence to stress, is analyzed from these measurements.
The stress-optical coefficient under uniaxial extensional flow, obtained using the proposed technique, is confirmed to be comparable to that under shear flow and to depend on the number of oriented micelles. 
\end{abstract}

\maketitle

\section{\label{sec:level1}Introduction}
A rheo-optical technique \cite{philippoff1957flow, chow1984response, meyer1993investigation, humbert1998flow, sridhar2000birefringence, haward2011extensional, haward2012extensional, pathak2006rheo, humbert1998stress, decruppe2003flow, ge2012rheo} involving simultaneous rheological and optical measurements is suitable for investigating the relationship between the rheology of complex fluids and their internal structural deformation under stress loading.
One of the optical measurements used in the rheo-optical technique is birefringence measurement,\cite{bass2010handbook, rastogi2015digital} and recently, the use of high-speed polarization cameras has enabled two-dimensional measurement of birefringence fields in complex fluids under shear flow conditions.\cite{ito2015temporal,ito2016shear,worby2024examination,yamamoto2024elucidating,sato2024two}
Quantifying changes in the optical properties of a component using birefringence measurements enables the estimation of the orientation state of the microstructure inside complex fluids. 
By combining the macro- and microscale results from viscoelasticity and birefringence measurements, respectively, the structural deformation inside complex fluids under stress loading can be comprehensively evaluated. 
The impact of the present technique is derived by implementing the quality of optical products, such as pickup lenses, optical discs, and liquid crystal films. 
For example, in the injection molding method \cite{mckelvey1962polymer,angstadt2006investigation, kim2005modeling, min2011experimental, lin2019grey} for manufacturing optical products, the molten material is exposed to strong shear and extensional flows during the injection molding, inducing residual stress. 
Because birefringence is proportional to the principal stress difference, the residual stress induces birefringence, which affects the quality of the optical products, causing issues, such as optical aberrations and color irregularities.\cite{lin2019grey} 
Therefore, product quality can be improved by understanding the stress loading state during injection molding and controlling the induced birefringence.

The proportional relationship between birefringence and the principal stress difference is known as the stress-optical rule.\cite{noto2020applicability,rastogi2003photomechanics,rastogi2013optical} %ryu1996simple,inoue1991birefringence
By applying stress to the component that exhibits birefringence, the birefringence $\delta n$ [–] for the principal stress difference $\sigma_{\parallel}-\sigma_{\perp}$ [Pa] can be expressed as 
\begin{equation}
\delta n = n_{\parallel}-n_{\perp} = C (\sigma_{\parallel}-\sigma_{\perp})
\label{eq1},
\end{equation}
where $n_{\parallel}$ [–] and $n_{\perp}$ [–] denote the refractive indices for polarizations parallel and perpendicular to the oriented birefringent objects, such as polymer chains and micelles, respectively.
The principal stress difference is the difference between principal stresses perpendicular to each other. 
The principal stresses ($\sigma_{\parallel}$ and $\sigma_{\perp}$) are proportional to the refractive indices ($n_{\parallel}$ and $n_{\perp}$) corresponding to each principal stress direction.
The proportionality coefficient $C$ [$\rm Pa^{-1}$] is called the stress-optical coefficient and indicates the sensitivity of the birefringence of the component to stress.\cite{noto2020applicability,rastogi2003photomechanics,rastogi2013optical} 
A general birefringence measurement technique enables the optical measurement of the phase retardation $\delta$ [nm] as the integrated value of birefringence $\delta n$, expressed as follows:
\begin{equation}
\delta = \int \delta n\,dz= C \int (\sigma_{\parallel}-\sigma_{\perp})\,dz
\label{eq2},
\end{equation}
where the $z$-axis direction represents the direction of light travel. 
The measured data of the phase retardation $\delta$ and stress-optical coefficient $C$ of the component must be obtained to investigate the stress applied inside the component.

To validate the stress-optical rule, it is necessary to demonstrate the intrinsic physical properties of the stress-optical coefficient.
For solids, several studies have investigated\cite{rastogi2013optical,rastogi2003photomechanics,ebisawa2007mechanical,sun2023measurement} to obtain the photoelastic coefficient using the photoelastic method. 
For fluids, on the other hand, some studies have investigated flow birefringence measurement of complex fluids, and measured the stress-optical coefficients under shear flow.\cite{talbott1979streaming,humbert1998flow,doyle1998relaxation,sridhar2000birefringence,janeschitz2012polymer,haward2011extensional,haward2012extensional,zhao2016flow,iwata2019local,kadar2021cellulose,lane2022birefringent} 
Micellar solutions are typical examples of viscoelastic fluids that induce birefringence, and their birefringence fields have been widely measured.\cite{shikata1994rheo,wheeler1996structure,C_Takahashi,ito2015temporal,ito2016shear,haward2019flow,cardiel2016formation,ge2012rheo,decruppe2003flow,humbert1998stress} 
Shikata et al. \cite{shikata1987micelle,shikata1988micelle,shikata1989micelle,shikata1988nonlinear} reported that micellar solutions of cetyltrimethylammonium bromide (CTAB) and sodium salicylate (NaSal) form extremely long micelles at relatively low concentrations and exhibit remarkable viscoelasticity.
In addition, CTAB/NaSal solutions exhibited birefringence under stress loading.
To measure the stress-optical coefficient of CTAB/NaSal solutions under shear flow, some researchers have used a rheo-optical technique with a coaxial cylinder rheometer or parallel plate rheometer.\cite{shikata1994rheo,wheeler1996structure,C_Takahashi,humbert1998flow,ito2015temporal,ito2016shear}
%Ito et al. \cite{ito2015temporal,ito2016shear} confirmed that the birefringence of micellar solutions under shear flow induced by a coaxial cylinder rheometer is synchronized with its stress fluctuation and generates a shear-induced structure.

\begin{figure}[b!]
\includegraphics[width=\columnwidth]{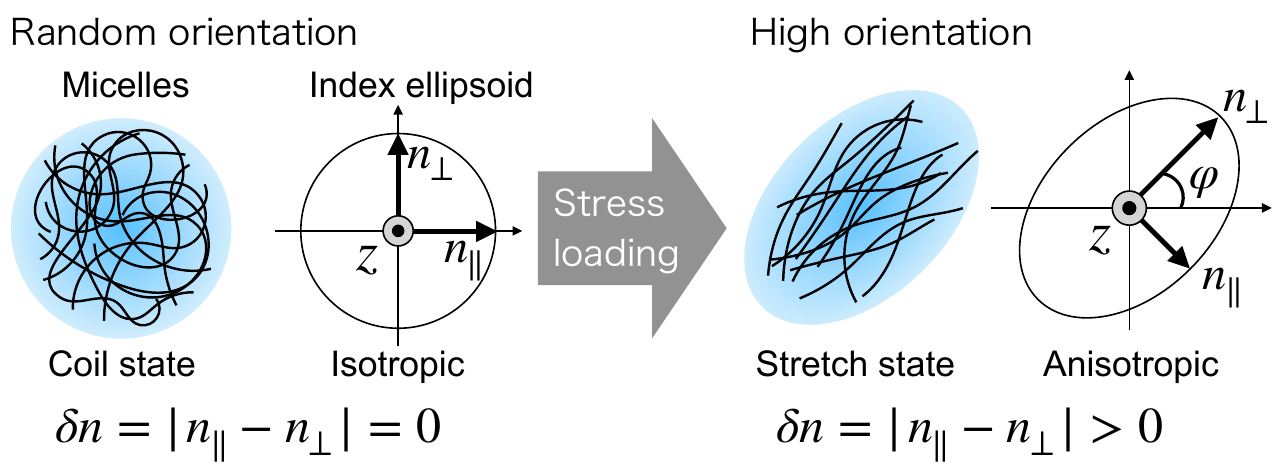}
\centering
\caption{\label{fig1} Mechanism of emergence of birefringence through micelle orientation under applied stress: When stress is applied to a micellar solution, the micelles extend and align at an orientation angle $\varphi$, resulting in increased orientational order. This alignment enhances the birefringence $\delta n$.}
\end{figure}

The birefringence of micelles can be attributed to the optical anisotropy induced by the index ellipsoid deformation,\cite{lodge1955variation,bass2010handbook,janeschitz2012polymer,decruppe2003flow} as shown in Fig.~\ref{fig1}. 
%===============================================
%===============================================
Without stress loading, the micelles in the solution had random orientations in the coiled state; thus, flow birefringence did not occur ($\delta n = |n_{\parallel}-n_{\perp}| = 0$).
When applying stress to the solutions, the micelles extend, align at the orientation angle $\varphi$ [rad], and become highly oriented, increasing birefringence ($\delta n = |n_{\parallel}-n_{\perp}| > 0$). 
Note that the emergence of birefringence does not require perfect alignment; even a Gaussian random orientation distribution can produce measurable birefringence owing to the ensemble-averaged optical anisotropy of the system.\cite{decruppe2003flow,shikata1994rheo}

Although flow birefringence measurements under extensional and shear flows are important for understanding the structural deformation of complex fluids, only a few studies have explored this phenomenon under an extensional flow. 
Pathak and Hudson \cite{pathak2006rheo} confirmed the birefringence of micellar solutions under planar extensional flow using a cross-slot channel.
Haward et al. \cite{haward2011extensional} used cross-slot channels to confirm the orientation of atactic polystyrene under a planar extensional flow. 
Rothstein et al.,\cite{rothstein2002comparison,rothstein2002inhomogeneous} Sridhar et al.,\cite{sridhar2000birefringence} and Talbott et al.\cite{talbott1979streaming} measured the birefringence of uniaxially extended filaments of polystyrene solutions using a filament-stretching rheometer. 
Furthermore, Rothstein \cite{rothstein2003transient} showed that the stress-optical rule breaks down in cases where wormlike micelles are significantly stretched and in a fully oriented state.

In the present study, a “full-field extensional rheo-optical technique” was developed to simultaneously measure the viscoelasticity and birefringence of micellar solutions under uniaxial extensional flow. 
The present technique integrates the liquid dripping method \cite{muto2023rheological, tamano2017dynamics, amarouchene2001inhibition} with a high-speed polarization camera, enabling the visualization of the full-field two-dimensional (2D) flow birefringence in complex fluids with high spatiotemporal resolution and the determination of their stress-optical coefficients.
In the liquid-dripping method, the extensional stress acting on the liquid filament is estimated without physical contact by analyzing the thinning dynamics of the filament as it undergoes uniaxial extension. 
By capturing synchronized high-speed images of the viscoelastic deformation of the filament and its full-field 2D flow birefringence, the birefringence corresponding to extensional stress can be quantitatively assessed. 
Using the present technique, we conducted rheo-optical measurements of CTAB/NaSal micellar solutions in uniaxial extensional flow. 
From the measured birefringence and extensional stress data, we evaluated the stress-optical coefficients of the micellar solutions by applying a simplified form of the stress-optical rule, as presented later in Eq.~(\ref{eq4}).

\section{\label{method}Experiments and methods}
\begin{figure*}[t]
\centering
\includegraphics[width=0.64\textwidth]{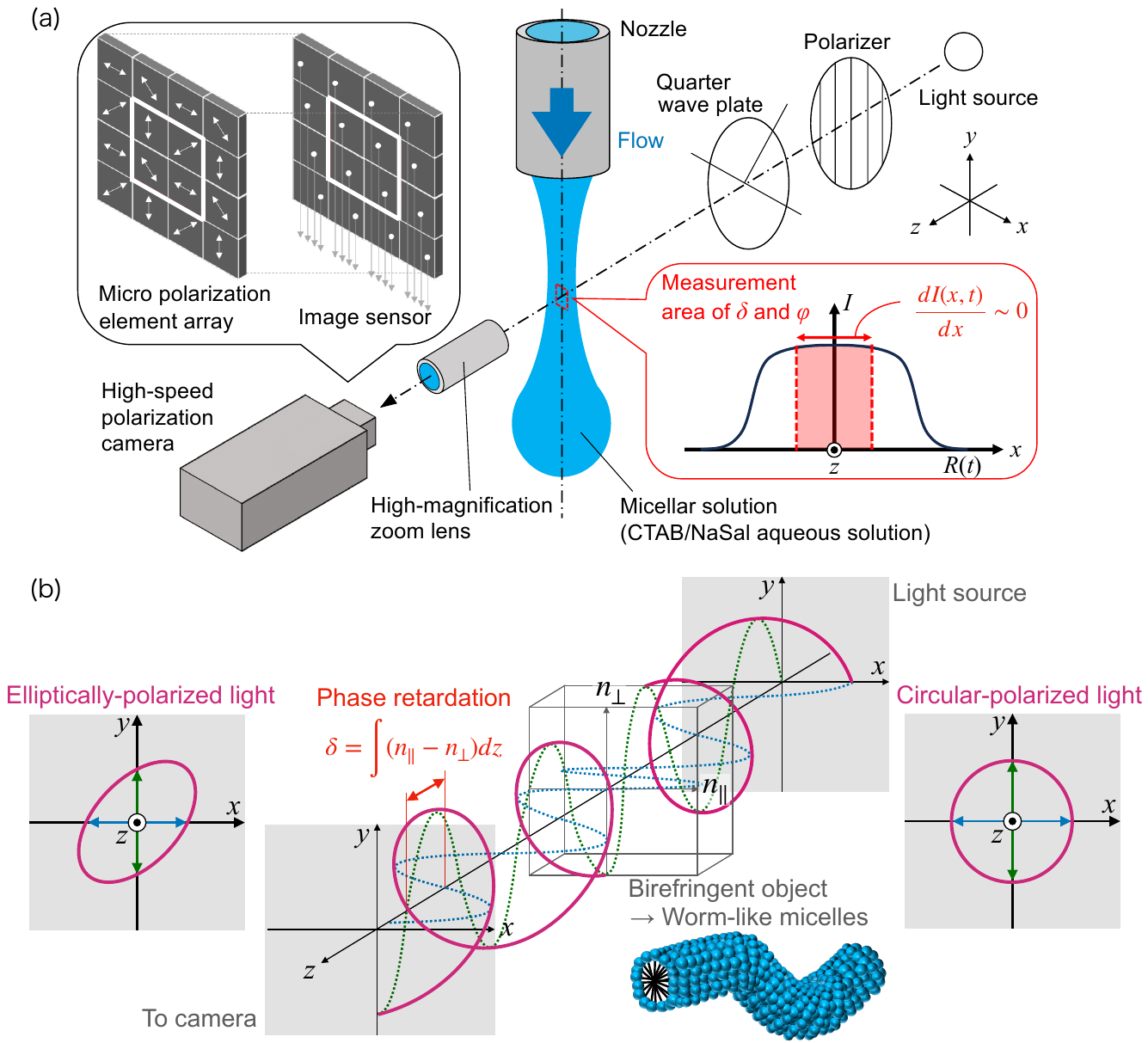}
\caption{\label{rheo_optical_technique}
Schematics of (a) the experimental setup for the full-field extensional rheo-optical technique and (b) the measurement principle of phase retardation in a birefringent micellar solution.
In Fig.~(a), the setup combines a high-speed polarization camera with a liquid-dripping method. The measurement area is located at the filament’s minimum diameter $R(t)$, where lensing effects are negligible and the radial light intensity profile remains uniform ($dI(x,t)/dx \sim 0$), indicating straight light propagation with minimal refraction.
In Fig.~(b), the blue and green curves represent the two orthogonal electric field components that make up the incident electromagnetic wave (red curve). The circular-polarized light emitted from the light source is transformed into elliptically-polarized light upon passing through the birefringent sample (micellar solution), and the resulting phase retardation $\delta$ is measured using the high-speed polarization camera.}
\end{figure*}

This section describes the measurement principles and advantages of the full-field extensional rheo-optical technique, which integrates flow birefringence measurements using a high-speed polarization camera with a liquid-dripping method, as illustrated in Fig.~\ref{rheo_optical_technique}(a). 
%===============================================
%===============================================
The specifications of the micellar solutions used as target samples in the present study are also provided.

\subsection{\label{rheo-optical technique}Flow birefringence measurement with high-speed polarization camera}
The concept of full-field extensional rheo-optical techniques can be realized using a simple setup: a sample placed between an LED light source (SOLIS-525C, Thorlabs, Inc.) and high-speed polarization camera (CRYSTA PI-1P, Photron Ltd.) equipped with a high-magnification zoom lens (Leica Z16APO, Leica Microsystems, Inc.).
The micellar solution used as the sample was extended uniaxially using the liquid-dripping method.
The birefringence measurements were employed to determine the phase retardation of polarized light transmitted through the sample under extensional stress loading.
Figure~\ref{rheo_optical_technique}(b) shows the changes in the polarization state of light when transmitted through a birefringent object.
The incident electromagnetic wave, represented by the red curve, has two orthogonal components, represented by the blue and green curves.
When the retardation of $\pi/2$ rad exists between the two components, the incident light is circularly polarized.
When light is transmitted through the birefringent object under stress loading, the propagation speed of each component wave differs because the refractive indices $n_{\parallel}$ and $n_{\perp}$ differ.
Thus, the phases of the components differed, and the retardation values changed.

The phase retardation $\delta$ is given by the integral of the birefringence $\delta n$ over infinitesimal volume elements within the birefringent object along the optical axis in the $z$-direction (refer to Eq.~(\ref{eq2})).
The high-speed polarization camera captures photoelastic phenomena at a frame rate of up to $1.5 \times 10^6$ fps.
The array consists of four adjacent polarizers ($2\times2$) arranged in four orientations.
The instantaneous retardation ${\delta}$ and orientation angle $\varphi$ were calculated from the light intensities captured by the four polarizers ($I_1$ at 0$^\circ$, $I_2$ at 45$^\circ$, $I_3$ at 90$^\circ$, and $I_4$ at 135$^\circ$).\cite{onuma2014development}
\begin{align}
{\delta} &= \frac{\lambda}{2 \pi} \sin ^{ -1 }{ \frac { \sqrt { { ( { I }_{ 3 }-{ I }_{ 1 } ) }^{ 2 }+{ ( { I }_{ 2 }{-I }_{ 4 } ) }^{ 2 } } }{ ({ I }_{ 1 }+{ I }_{ 2 }+{ I }_{ 3 }+{ I }_{ 4 } )/2} }
\label{delta},
\\[6pt]
\varphi &=\frac{1}{2}\tan ^{ -1 }{ \frac { { I }_{ 3 }-{ I }_{ 1 } }{{ I }_{ 2 }-{ I }_{ 4 }}}
\label{phi},
\end{align}
where $\lambda$ [nm] denotes the wavelength of the incident light.
In the present study, the incident light emitted from a green LED with a wavelength of 525 nm was polarized using a circular polarization sheet consisting of a polarizer and quarter-wave plate. 
The incident light was directed straight into the camera.
The orientation angle $\varphi$ is defined as the tilt angle of the micelles with respect to the horizontal ($x$) direction; the vertical ($y$) direction is the extension direction.

Table \ref{Table:camera} lists the shooting conditions for the present technique. 
%===================================================
\begin{table}[b!]
    \centering
    \caption{Shooting conditions.}
    \label{Table:camera}
    \renewcommand \arraystretch{0.9}
    \begin{tabular}{lc}
        \hline\hline
        Parameter & Specification \\ \hline
        Frame rate [fps] & 60 -- 2000 \\       
        Shutter speed [s] & 1/60 -- 1/2000 \\ 
        Magnification & 2 -- 4 \\
        Spatial resolution [µm/pixel] & 5 -- 10 \\
        Shooting area [mm$^2$] & 5.12 $\times$ 5.12 -- 10.24 $\times$ 10.24  \\
        (Shooting area [pixels]) & (1024 $\times$ 1024)  \\
        Measurement area [µm$^2$] & 20 $\times$ 20 -- 40 $\times$ 40\\
        \hline\hline
    \end{tabular}
\end{table}
%===================================================
To determine these conditions, it is necessary to observe the temporal evolutionary extension behavior of the filament and measure the phase retardation $\delta$ and orientation angle $\varphi$ at the filament center with high spatiotemporal resolutions. 
The shooting area and the measurement area are defined separately; the shooting area refers to the field of view capturing the entire liquid filament in either the light intensity images or the phase retardation images, whereas the measurement area refers to the localized region used to evaluate the $\delta$ and $\varphi$ at the point of stress concentration within the filament.
Because the filament has a cylindrical cross section, a lensing effect caused by refraction occurs when polarized light from the light source passes through its curved interface during side observation, preventing accurate measurement.
Therefore, the measurement area at the center of the filament, where the lensing effect was negligible, was selected, as shown in Fig.~\ref{rheo_optical_technique}(a).
In this area, the light intensity profile along the radial direction remains spatially uniform ($dI(x,t)/dx \sim 0$), and thus the light is presumed to travel along a straight path without significant refraction.
Note that the measurement area occupies only about several $0.001\%$ of the shooting area, indicating that it is an extremely small region.
The details for determining the measurement area are described in Section I\hspace{-1.2pt}I\hspace{-1.2pt}I. B.

\subsection{\label{liquid dripping method}Liquid dripping method}
The experimental setup for the liquid-dripping method was based on that used in our previous study.\cite{muto2023rheological}
A fluid-dispensing device with a syringe pump (Pump 11 Elite, Harvard Apparatus Ltd.) delivered a droplet of the micellar solution at a relatively low flow rate.
A liquid filament formed as the droplet slowly dripped from the nozzle with an outer diameter $2R_0$ = 1.49 mm.
By measuring the minimum necking radius $R$ of the filament, which decreases as it extends, the extensional stress, viscosity, strain, strain rate, and relaxation time can be calculated.\cite{muto2023rheological,tamano2017dynamics,amarouchene2001inhibition}
In the present technique, the same mathematical framework used in dripping-on-substrate capillary breakup extensional rheometry (CaBER-DoS) \cite{dinic2020flexibility, dinic2017pinch,  mckinley2000extract, clasen2006dilute, dinic2019macromolecular, dinic2015extensional, sur2018drop, mckinley2005visco, omidvar2019detecting} is applicable for deriving the rheological properties.

\begin{figure}[t!]
\includegraphics[width=0.84\columnwidth]{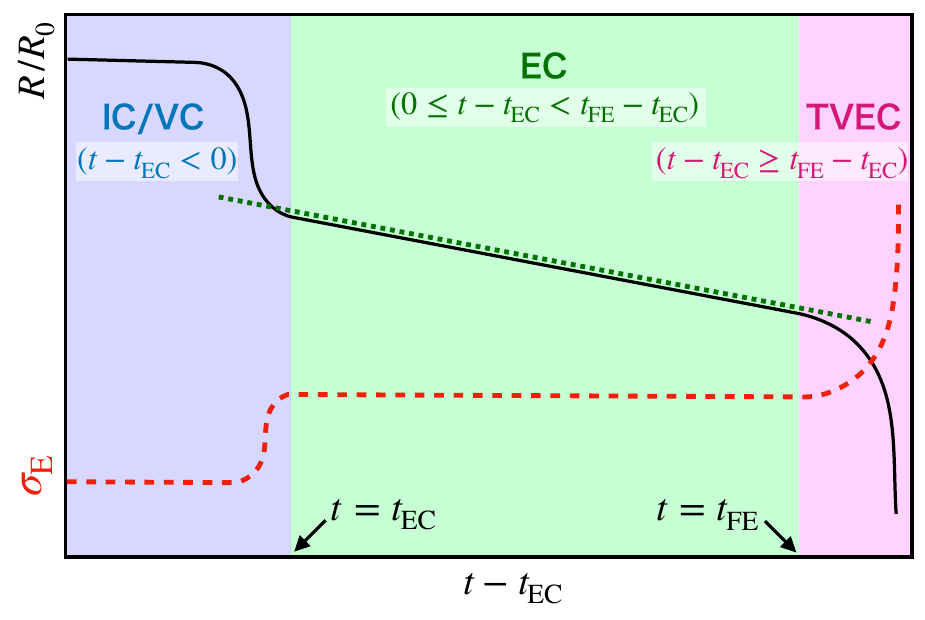}
\centering
\caption{Schematic of temporal evolution of radius ratio of liquid filament $R/R_0$ and extensional stress $\sigma_{\rm \hspace{1pt} E}$ for viscoelastic fluids. Note that the graph is a semi-logarithmic plot with a logarithmic scale on the vertical axis. Here, $t$, $t_{\mathrm{\hspace{1pt} EC}}$, and $t_{\mathrm{\hspace{1pt} FE}}$ represent the time, onset time of the EC regime, and onset time of the TVEC regime, respectively.
}
\label{EC_regime}
\end{figure}

To derive these physical quantities of viscoelasticity, the measured data for the radius ratio $R/R_{0}$ of the liquid filament within the elasto-capillary (EC) regime must be extracted, as shown in Fig.~\ref{EC_regime}.
%===================================================
%===================================================
Here, $R_{0}$ denotes the initial radius at the initial time ($t = 0$ ms): the outer radius of the nozzle in the present study.
The horizontal axis represents $t-t_{\mathrm{\hspace{1pt} EC}}$, where $t_{\mathrm{\hspace{1pt} EC}}$ [ms] denotes the onset time of the EC regime. 
The time evolution of the radius ratio $R/R_{0}$ can be classified into three regimes:\cite{dinic2019macromolecular} inertio-capillary/visco-capillary (IC/VC), EC, and terminal visco-elasto-capillary (TVEC).
The phase before the EC regime ($t - t_{\mathrm{\hspace{1pt} EC}} < 0$) is known as the IC/VC regime, whereas Newtonian fluids exhibit only the IC/VC regime. 
In contrast, viscoelastic fluids exhibit an EC regime. 
By defining the onset time of the TVEC regime as $t_{\mathrm{FE}}$ [ms], the EC regime can be defined as the interval $0 \leq t - t_\mathrm{\hspace{1pt} EC} < t_\mathrm{\hspace{1pt} FE} - t_\mathrm{\hspace{1pt} EC}$. 
In the duration extending beyond the EC regime up to the point just before the filament breaks ($t - t_{\mathrm{\hspace{1pt} EC}} \geq t_{\mathrm{\hspace{1pt} FE}} - t_{\mathrm{\hspace{1pt} EC}}$), the TVEC regime can be identified by the rapid reduction in the radius ratio of the filament over time.

Theoretically, the EC regime is a pure uniaxial extension with exponential decay of the radius ratio $R/R_{0}$ over time $t - t_{\mathrm{\hspace{1pt} EC}}$, expressed as follows:\cite{dinic2019macromolecular,dinic2015extensional,sur2018drop,mckinley2005visco,wagner2015analytic,prabhakar2006effect,omidvar2019detecting}
\begin{equation}
\frac{R(t)}{R_{0}} = \left(\frac{GR_{0}}{2\Gamma}\right)^{1/3}\exp\!\left(-\frac{t - t_{\mathrm{\hspace{1pt} EC}}}{3\lambda_\mathrm{E}}\right)
\label{radius in ECregime},
\end{equation}
where $G$ [Pa], $\Gamma$ [mN/m], and $\lambda_\mathrm{E}$ [ms] denote the elastic modulus, surface tension, and extensional relaxation time of the fluid, respectively. 
The onset time of the EC regime $t_\mathrm{\hspace{1pt} EC}$ is determined as the intersection point where the exponential thinning behavior of $\ln{R(t)/R_{0}}$ begins to follow a linear trend.

Within the EC regime, where the elastic tensile force of the fluid balances the capillary force, the filament of the viscoelastic fluid extends at an approximately constant strain rate.
Under this condition, the extensional stress $\sigma_{\rm \hspace{1pt} E}$ [Pa] can be reliably estimated from the capillary pressure as:\cite{Anna2001,McKinley2002}
\begin{equation}
\sigma_{\rm \hspace{1pt} E} = \eta_{\rm \hspace{1pt} E} \dot{\varepsilon} = \frac{\Gamma}{R(t)}
\label{stress in ECregime},
\end{equation}
where $\eta_{\rm \hspace{1pt} E}$ [$\mathrm{Pa} \cdot \mathrm{s}$] and $\dot{\varepsilon}$ [s$^{-1}$] denote the fluid extensional viscosity and extensional rate, respectively. 
Here, the extensional rate $\dot{\varepsilon}$ is determined from the temporal evolution of the filament diameter as:\cite{Anna2001,McKinley2002}
\begin{equation}
\dot{\varepsilon}(t) = -\frac{2}{R(t)} \frac{dR(t)}{dt}
\label{strainrate in ECregime}.
\end{equation}
The approximations expressed in Eqs.~\eqref{stress in ECregime} and \eqref{strainrate in ECregime} are widely accepted for slender filaments undergoing uniaxial extensional flow, where both inertial and viscous effects are negligible.\cite{Anna2001,McKinley2002}
Within the EC regime, the extensional stress varies only slowly over time compared to other regimes, and is often treated as quasi-steady.

\subsection{\label{Advantages of full-field extensional rheo-optical technique}Advantages of full-field extensional rheo-optical technique}
In the full-field extensional rheo-optical technique, uniaxial extensional flow aligns the micelles in the extensional direction of the solution, offering the following three advantages.

\begin{figure}[tb!]
\includegraphics[width=0.9\columnwidth]{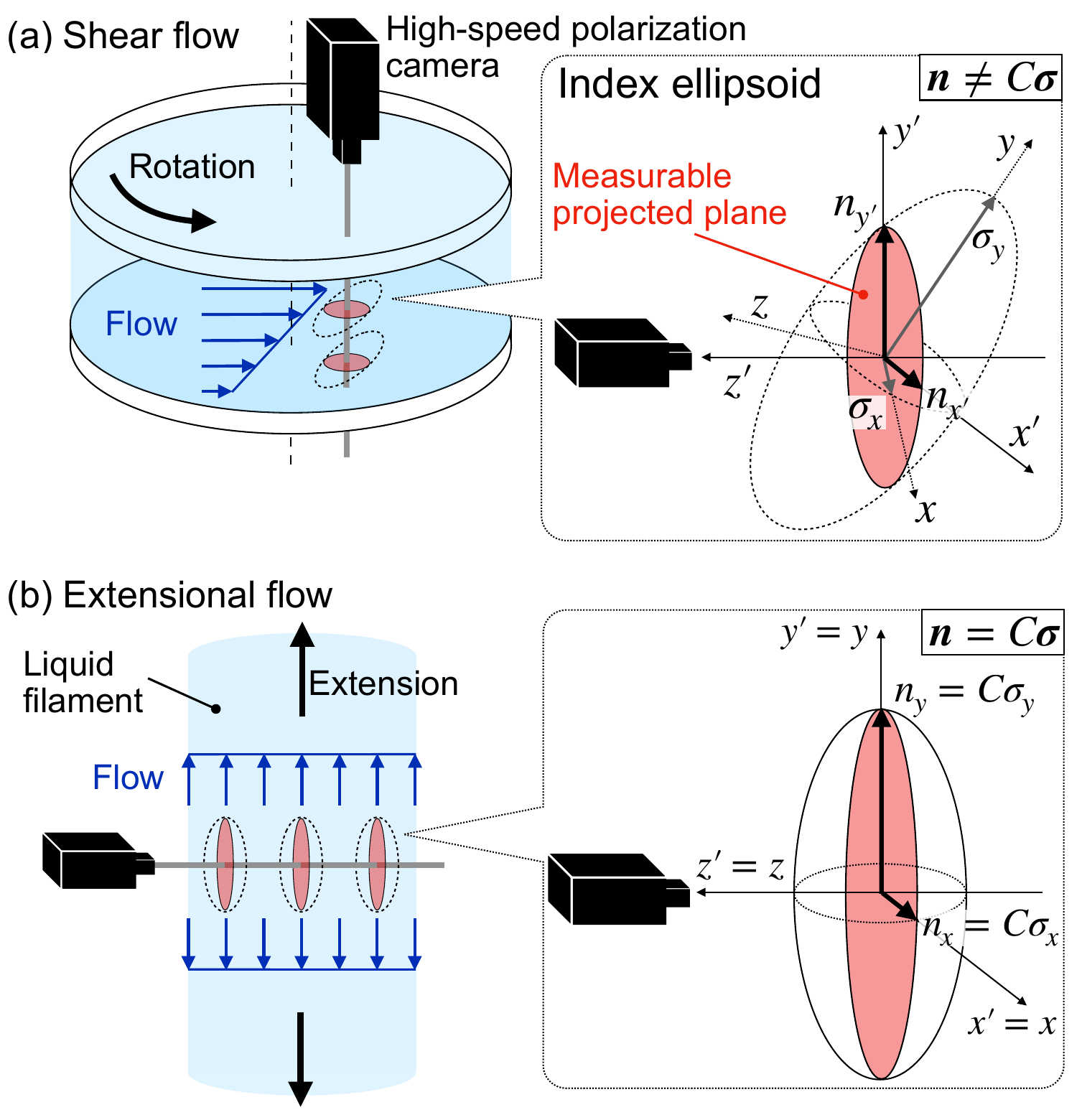}
\centering
\caption{\label{fig2} 
Conditions of index ellipsoids under (a) simple shear and (b) uniaxial extensional flows. 
Here, the major axis ($n_{\perp}$) and minor axis ($n_{\parallel}$) of refractive indices in the index ellipsoid correspond to $n_{y}$ and $n_{x}$, respectively. $x, y, z$ denote the coordinate system based on the refractive index ellipsoid, and $x', y', z'$ denote the coordinate system aligned with the optical axis of the high-speed polarization camera. In case (b), both coordinate systems coincide.}
\end{figure}

First, the major axis of the refractive index in the index ellipsoid was oriented in the uniaxial extensional direction; thus, the stress-optical rule was valid.
The birefringence measurement in the rheo-optical technique allows the measurement of the differences between the major axis ($n_{\perp}$) and minor axis ($n_{\parallel}$) of the refractive indices in the plane perpendicular to the optical axis.
However, in a simple shear flow, the index ellipsoid is tilted with respect to the optical axis owing to the complex orientation of micelles along an intermediate direction between the flow and velocity gradient axes (Fig.~\ref{fig2}(a)).
%===============================================
%===============================================
Consequently, birefringence results from a mixed projection of the refractive-index tensor, and the measured value no longer directly reflects the principal-stress direction.
This geometric misalignment results in a breakdown of the stress-optical rule because the refractive index vector $\bm{n}$ no longer scales directly with the principal stress tensor $\bm{\sigma}$, that is, $\bm{n} \neq C \bm{\sigma}$.
Consequently, it becomes difficult to extract accurate stress information from birefringence data under simple shear flow.
However, birefringence measurements under uniaxial extensional flow using the liquid-dripping method avoided this problem (Fig.~\ref{fig2}(b)).
In this case, the index ellipsoid was axisymmetric and aligned with the optical geometry, with its major axis oriented along the extensional direction and perpendicular to the optical axis.
This geometrical alignment allows the measured birefringence to correspond directly to the principal stress component, thereby restoring the validity of the stress-optical rule, $\bm{n} = C \bm{\sigma}$.
Therefore, accurate and direction-resolved measurements of birefringence are enabled, facilitating robust determination of the stress-optical coefficient under uniaxial extensional flow.

Second, only the principal stress in one direction affects the uniaxially extended liquid filament, thus simplifying Eq.~(\ref{eq2}) for the principal stress difference.  
Because there is no flow in the radial direction of the filament, the principal stress in the horizontal component can be neglected ($\sigma_{\parallel} = 0$).  
This is a significant simplification compared with shear flow, where multiple stress components and their spatial gradients contribute to the optical response.  
Furthermore, the principal stress in the vertical component is rewritten as uniaxial extensional stress ($\sigma_{\perp} = \sigma_{\rm E}$), which can be directly calculated by measuring the minimum radius of the filament using the capillary balance relationship (refer to Eq.~(\ref{stress in ECregime})) within the EC regime.\cite{Anna2001,McKinley2002}  
By observing the lateral extension of the liquid filament, the optical path length traversed by the polarized light can be considered equal to the filament diameter ($2R$), under the assumption of cylindrical symmetry.
Consequently, the phase retardation $\delta$ arises solely from the stress-induced optical anisotropy along the direction of the principal stress difference, thereby reducing Eq.~(\ref{eq2}) to:
\begin{equation}
\delta =\delta n \cdot 2R=C \sigma_{\rm E}\cdot 2R.
\label{eq3}
\end{equation}
This expression results in a direct evaluation of the stress-optical coefficient.
\begin{equation}
C = \frac{\delta n}{\sigma_{\rm E}}.
\label{eq4}
\end{equation}
These relations are valid only when the incident polarized light passes through the center of the filament, where both the flow field and stress distribution exhibit radial symmetry, and the optical path is well defined.

Third, employing a high-speed polarization camera enables full-field 2D birefringence measurements with high spatiotemporal resolution, providing a distinct advantage over conventional pointwise laser-based techniques in capturing spatiotemporally varying stress concentration areas.  
In a uniaxial extensional flow, the liquid filament may exhibit a slight curvature along the stretching direction owing to gravity or surface tension, resulting in nonuniform distributions of extensional stress along the filament axis~\cite{clasen2006dilute}.
The full-field 2D imaging enables accurate and high-speed identification of the localized maximum stress region near the necking point, where the filament diameter is minimal, as shown in Table \ref{Table:camera}.
Furthermore, this approach also minimizes the influence of the lensing effect, which arises from light refraction due to the radial curvature of the filament (as discussed in detail later).
This eliminates the spatial alignment sensitivity inherent to pointwise methods, thereby enhancing the spatial resolution and reliability of the birefringence–stress correspondence.

\subsection{\label{micellar soludions}Micellar solutions}
In the present study, 17 micellar solutions with different molar concentrations of CTAB (FUJIFILM Wako Pure Chemical Co.) and NaSal (Sigma-Aldrich Co.) were used, as listed in Table \ref{Table:solutions}.
%===============================================
\begin{table}[tb!]
    \centering
    \caption{CTAB/NaSal aqueous solutions with varied concentrations of CTAB ($c_\mathrm{D}$) and NaSal ($c_\mathrm{S}$) used in the present study. $c_\mathrm{V}$ and $c_\mathrm{S}/c_\mathrm{D}$ indicate micelle molar concentration and CTAB/NaSal molar ratio, respectively.}
    \label{Table:solutions}
    \tabcolsep = 0.2cm
    \renewcommand \arraystretch{1.35}
    \begin{tabular}{c c c c c}
    \hline \hline
        Solution & $c_\mathrm{D}$ & $c_\mathrm{S}$ & $c_\mathrm{V}$ & $c_\mathrm{S}/c_\mathrm{D}$ \\ 
        No. & [M] & [M] & ($\sim c_\mathrm{D}$)[M] & [–] \\ \hline 
        $\mathrm I$ &  0.005  & 0.038 & 0.005 & \multirow{5}{*}{7.7} \\ 
        $\mathrm {I\hspace{-0.6pt}I}$ &  0.010  & 0.077 & 0.010 &  \\ 
        $\mathrm {I\hspace{-0.6pt}I\hspace{-0.6pt}I}$ &  0.015  & 0.115 & 0.015 &  \\
        $\mathrm {I\hspace{-0.6pt}V}$ &  0.020  & 0.153 & 0.020 & \\
        $\mathrm V$ &  0.025  & 0.192 & 0.025 & \\
        \hline
        $\mathrm {V\hspace{-0.6pt}I}$ &  0.030  & 0.230 & 0.030 & 7.7 \\
        \hline
        $\mathrm {V\hspace{-0.6pt}I\hspace{-0.6pt}I}$ &  \multirow{11}{*}{0.030}  & 0.030 & \multirow{11}{*}{0.030} & 1.0 \\ 
        $\mathrm {V\hspace{-0.6pt}I\hspace{-0.6pt}I \hspace{-0.6pt} I}$ &    & 0.060 & & 2.0 \\
        $\mathrm {I\hspace{-0.6pt}X}$ &    & 0.090 & & 3.0 \\ 
        $\mathrm {X}$ &    & 0.120 &  & 4.0 \\
        $\mathrm {X\hspace{-0.6pt}I}$ &    & 0.150 &  & 5.0 \\
        $\mathrm {X\hspace{-0.6pt}I\hspace{-0.6pt}I}$ &    & 0.180 &  & 6.0 \\
        $\mathrm {X\hspace{-0.6pt}I\hspace{-0.6pt}I\hspace{-0.6pt}I}$ &  & 0.210 &  & 7.0 \\
        $\mathrm {X\hspace{-0.6pt}I\hspace{-0.6pt}V}$ &    & 0.270 &  & 9.0 \\
        $\mathrm {X\hspace{-0.6pt}V}$ &   & 0.300 &  & 10 \\
        $\mathrm {X\hspace{-0.6pt}V\hspace{-0.6pt}I}$ &    & 0.330 &  & 11 \\
        $\mathrm {X\hspace{-0.6pt}V\hspace{-0.6pt}I\hspace{-0.6pt}I}$ &    & 0.360 &  & 12 \\
        \hline \hline
    \end{tabular}
\end{table}
%===============================================
The molar concentrations of CTAB and NaSal are denoted $c_\mathrm{D}$ [M] and $c_\mathrm{S}$ [M], respectively. 
Moreover, $c_\mathrm{V}$ [M] represents the micelle molar concentration, which is defined as the ratio of the total micelle volume to the entire solution volume and indicates the total amount of micelles in the solution.
Shikata et al.\cite{shikata1987micelle,shikata1988micelle} experimentally demonstrated that micelles in CTAB/NaSal solutions were formed at a 1:1 molar ratio of NaSal to CTAB. 
Therefore, in solutions with $c_\mathrm{S}-c_\mathrm{D}>0$, $c_\mathrm{D}$ indicates the micelle molar concentration $c_\mathrm{V}$.
Previous studies on the rheo-optical measurements of micellar solutions under shear flow \cite{shikata1994rheo,wheeler1996structure,C_Takahashi,ito2015temporal} employed the conditions of $c_\mathrm{D}=0.030$ M and $c_\mathrm{S}=0.230$ M. 
Our study adopted similar conditions as the standard (solution $\mathrm {V\hspace{-0.6pt}I}$), prepared five solutions (solutions $\mathrm I$ to $\mathrm{V}$) with a fixed CTAB/NaSal molar ratio of $c_\mathrm{S}/c_\mathrm{D} = 7.7$, and eleven solutions (solutions $\mathrm {V\hspace{-0.6pt}I\hspace{-0.6pt}I}$ to $\mathrm {X\hspace{-0.6pt}V\hspace{-0.6pt}I \hspace{-0.6pt} I}$) with a fixed micelle molar concentration of $c_\mathrm{V}=0.030$ M, totaling seventeen solutions. 
The CTAB concentration $c_\mathrm{D}$ of all the solutions exceeded the critical micelle concentration of $8.9 \times 10^{-4}$~M.\cite{paredes1984enthalpies} 
Within the CTAB concentration range of 0.005 -- 0.030~M and a CTAB/NaSal molar ratio of 1.0 -- 12, micellar structures can be classified as either wormlike or networked, depending on the molar ratio. 
At the molar ratios of 1.0 -- 2.0, entangled wormlike micelles dominate. 
In contrast, when the molar ratio exceeds 3.0, particularly above 6.0, excess Sal$^-$ results in the formation of bridged micellar networks, which are entangled structures composed of wormlike micelles.\cite{kadoma1997structural}

\begin{figure}[tb!]
\includegraphics[width=0.96\columnwidth]{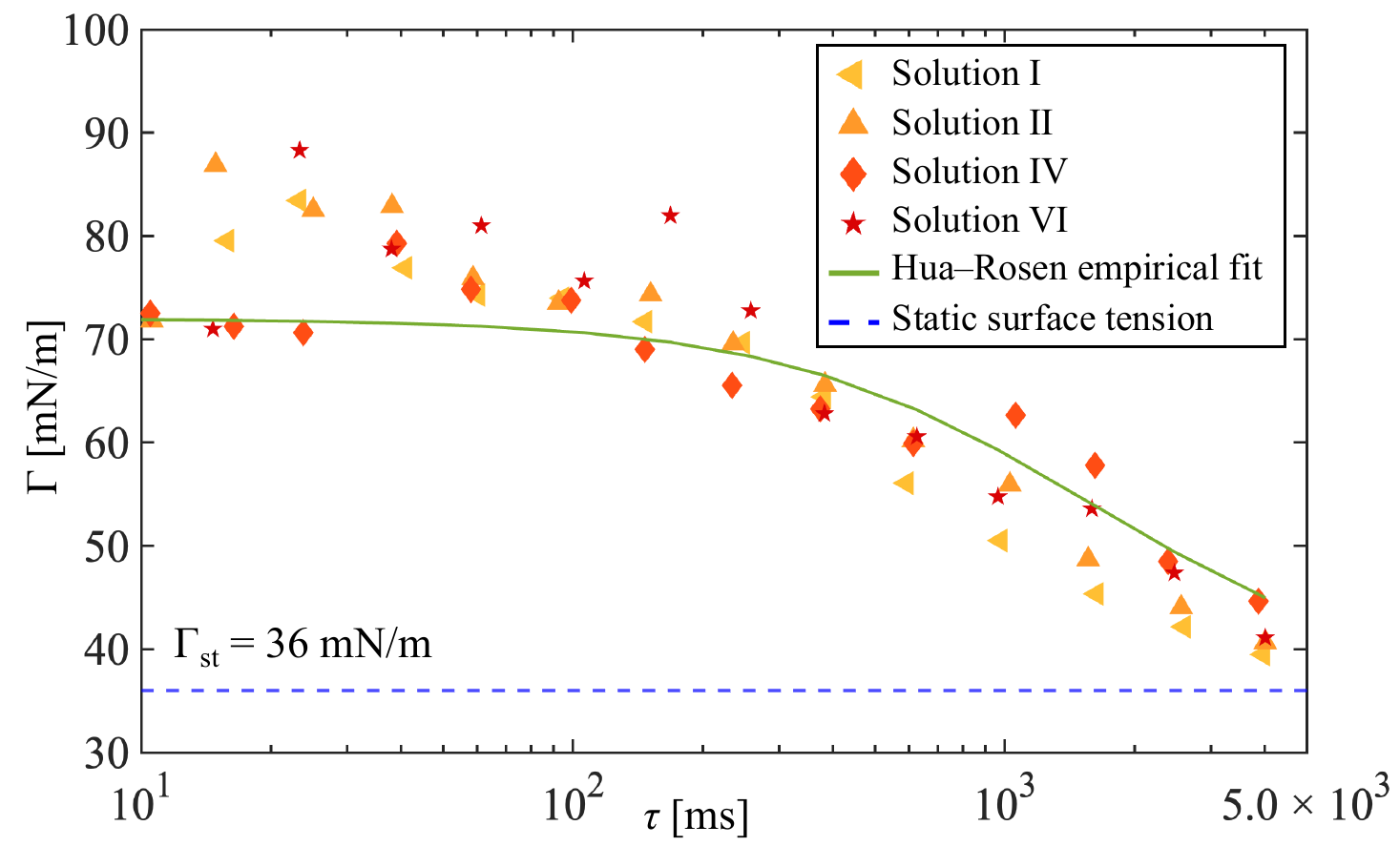}
\centering
\caption{Dynamic surface tension $\Gamma$ as a function of surface lifetime $\tau$ for representative micellar solutions (solutions $\mathrm{I}$, $\mathrm{II}$, $\mathrm{IV}$, $\mathrm{VI}$) at a constant CTAB/NaSal molar ratio ($c_\mathrm{S} / c_\mathrm{D} = 7.7$).
The curve in the graph represents an empirical fit based on the model proposed by Hua and Rosen,\cite{hua1988dynamic,rosen1990dynamic} applied to the averaged data from the solutions.}
\label{dynamic_surface_tension}
\end{figure}

For CTAB/NaSal solutions exceeding the critical micelle concentration, the static surface tension was $36$~mN/m.\cite{cooper2002drop}
This value was confirmed in the present study using a static surface tensiometer (DY-300, Kyowa Interface Science Co., Ltd.) and the Wilhelmy plate method.\cite{gaonkar1987uncertainty}
Conversely, because CTAB/NaSal solutions exhibit dynamic surface tension, the values of representative micellar solutions (solutions $\mathrm I$, $\mathrm {I\hspace{-0.6pt}I}$, $\mathrm {I\hspace{-0.6pt}V}$, $\mathrm {V\hspace{-0.6pt}I}$) at a constant CTAB/NaSal molar ratio ($c_\mathrm{S} / c_\mathrm{D} = 7.7$) were measured using a bubble pressure dynamic surface tensiometer (BP100, KRÜSS Scientific Instruments, Inc.). 
Approximately 100 mL of the sample was required for each measurement, which was conducted at room temperature (25~$^\circ$C) over a surface lifetime range of 10 -- 4000~ms.
Consequently, all solutions exhibited a decreasing trend in dynamic surface tension $\Gamma$ [mN/m] with increasing surface lifetime $\tau$ [ms], and their values were found to be of similar magnitude (Fig.~\ref{dynamic_surface_tension}). 
%===================================================
%===================================================
According to Cooper-White et al., \cite{cooper2002drop} the dynamic surface tension of CTAB/NaSal solutions is initially equivalent to that of pure water ($72~\mathrm{mN/m}$ at 25~$^\circ$C) within approximately the first 15 ms after new surface formation, and then gradually approaches an equilibrium value of $36~\mathrm{mN/m}$ at approximately 1000~ms. 
The trends observed in the present experimental results (Fig.~\ref{dynamic_surface_tension}) are consistent with the previous findings.\cite{cooper2002drop}

The empirical fit shown in Fig.~\ref{dynamic_surface_tension} was performed using the empirical equation proposed by Hua and Rosen,\cite{hua1988dynamic,rosen1990dynamic} and the resulting fit is also presented. 
The time-dependent surface tension $\Gamma(\tau)$ at the surface lifetime $\tau$ is expressed as follows:
\begin{equation}
\Gamma(\tau) = \Gamma_{\mathrm{st}} + \frac{(\Gamma_0 - \Gamma_{\mathrm{st}})}{1 + \left(\tau/\tau^* \right)^m},
\end{equation}
where $\Gamma_0$ [mN/m] denotes the surface tension of the pure solvent, $m$ [–] is an empirical constant, and $\tau^*$ [ms] represents the adsorption relaxation time. 
The static (equilibrium) surface tension measured using a static surface tensiometer was $\Gamma_{\mathrm{st}} = 36~\mathrm{mN/m}$.
Pure water was used as the solvent and the surface tension of the pure solvent was $\Gamma_0 = 72~\mathrm{mN/m}$.
The adsorption relaxation time was determined to be $\tau^* =$ 1600 ms, at which the surface tension reaches the midpoint value: $\Gamma(\tau^*) = (\Gamma_0 + \Gamma_{\mathrm{st}})/2 = 54$ mN/m. 

The liquid-dripping method was not significantly affected by the time-dependent surface tension characteristics of CTAB/NaSal solutions.  
This is because the Péclet number $\mathrm{Pe}$ [–] remains much greater than unity ($\mathrm{Pe} \gg 1$) throughout the extension process, from the onset of stretching to filament breakup.
The Péclet number is defined as the ratio between the diffusive time scale of micelles at the interface ($\tau_\mathrm{D}$ [s]) and the surface deformation time scale of the filament ($\tau_\mathrm{\Gamma}$ [s]) and is expressed as:\cite{regev2010role,brust2013rheology}
\begin{equation}
\mathrm{Pe} = \frac{\tau_\mathrm{D}}{\tau_\mathrm{\Gamma}} = \frac{R_{0} \, \Gamma}{4 D \, \eta_\mathrm{E}},
\label{eq:Pe}
\end{equation}
where $D$ [m$^2$/s] denotes the diffusion coefficient of the micelles, which has been reported to be approximately $10^{-11}$ -- $10^{-12}$ m$^2$/s.\cite{rehage1991viscoelastic} 
In addition, the extensional viscosity $\eta_\mathrm{E}$ observed in the present study was valid from $10^{0}$ to $10^{4}$ Pa$\cdot$s.
By applying these values, the order-of-magnitude estimation yields a Péclet number in the range of $10^2$ to $10^7$, confirming that $\mathrm{Pe} \gg 1$ is satisfied.  
Therefore, micelles were advected away by the bulk flow before they could be adsorbed at the interface, indicating that the influence of interfacial adsorption and time-dependent surface tension was negligible under the present experimental conditions.

\section{\label{result}Results and discussion}
This section reports the extensional stress and birefringence measurements obtained using the full-field extensional rheo-optical technique, along with the resulting evaluation of the stress-optical coefficient.
The section describing the derivation of the stress-optical coefficient is based on the results of the solutions $\mathrm {I}$ to $\mathrm {V\hspace{-0.6pt}I}$, in which the CTAB/NaSal molar ratio is kept constant. 
Conversely, the sections discussing the dependence of the stress-optical coefficient on the micelle molar concentration and CTAB/NaSal molar ratio are based on the results of all solutions (solutions $\mathrm {I}$ to $\mathrm {X\hspace{-0.6pt}V\hspace{-0.6pt}I \hspace{-0.6pt} I}$).
All experiments were conducted at room temperature of 25~$^\circ$C.

\subsection{\label{Extensional stress}Extensional behavior and stress}
\begin{figure*}[htbp!]
\includegraphics[width=0.82\textwidth]{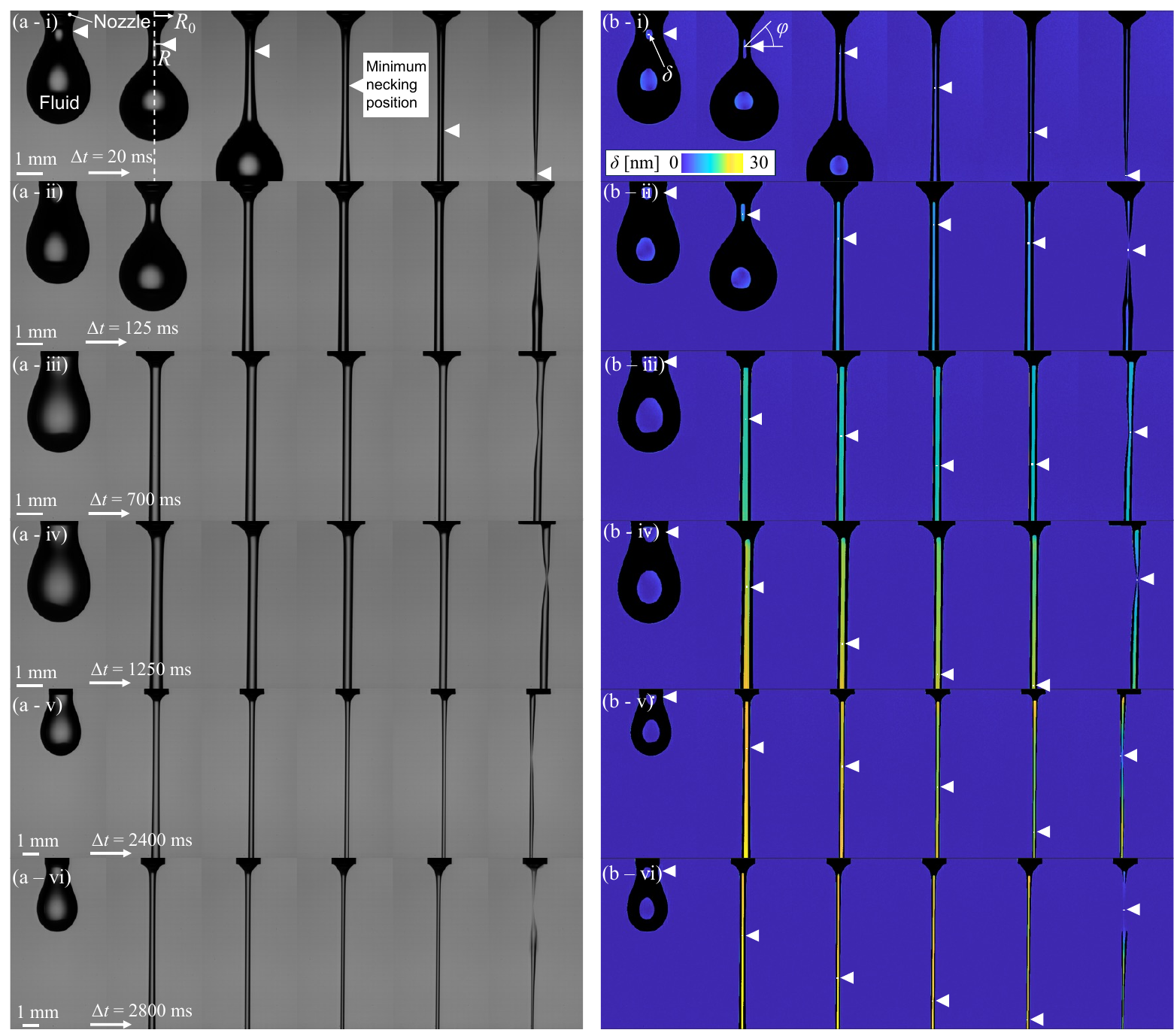}
\centering
\caption{Temporal images of light intensity and phase retardation of uniaxially extending micellar solutions at a constant CTAB/NaSal molar ratio ($c_\mathrm{S} / c_\mathrm{D} = 7.7$), both of which were simultaneously obtained in a single acquisition using the high-speed polarization camera. The left column contains (a) light intensity images, and the right column contains (b) phase retardation images, corresponding to different solutions: (i) solution $\mathrm I$, (ii) solution $\mathrm {I\hspace{-0.6pt}I}$, (iii) solution $\mathrm {I\hspace{-0.6pt}I\hspace{-0.6pt}I}$, (iv) solution $\mathrm {I\hspace{-0.6pt}V}$, (v) solution $\mathrm {V}$, (vi) solution $\mathrm {V\hspace{-0.6pt}I}$. $\Delta \hspace{0.4pt} t$ indicates time interval between each image. The white square indicates the measurement area at the filament center where the radius is minimal. The original images captured by the high-speed polarized camera had a resolution of 1024 $\times$ 1024 pixels. The images were cropped to 575 $\times$ 1024 pixels to exclude areas not showing the droplets. The deep blue background in the phase retardation images does not contribute to the contours because the orientation angle was random and the measured phase retardation was zero.}
\label{intensity_retardation_images}
\end{figure*}

Temporal images of the light intensity and phase retardation of CTAB/NaSal solutions $\mathrm I$ to $\mathrm {V\hspace{-0.6pt}I}$ at various concentrations while maintaining a constant CTAB/NaSal molar ratio ($c_\mathrm{S}/c_\mathrm{D}=7.7$) are shown in Fig.~\ref{intensity_retardation_images}, both of which were simultaneously obtained in a single acquisition using the high-speed polarization camera.
%===================================================
%===================================================
The minimum radius $R$ of the filaments, indicated by the white arrows in Fig.~\ref{intensity_retardation_images}(a-i), must be measured to investigate the extensional behavior of the filaments quantitatively. 
The temporal development of the extended liquid filament exhibits the thinning characteristics of viscoelastic fluids.
The time required for these extended filaments to break increases with increasing molar concentrations. 
Further discussion of the phase-retardation images (Fig.~\ref{intensity_retardation_images}(b)) is provided in Section I\hspace{-1.2pt}I\hspace{-1.2pt}I. B.

\begin{figure}[tb!]
\includegraphics[width=\columnwidth]{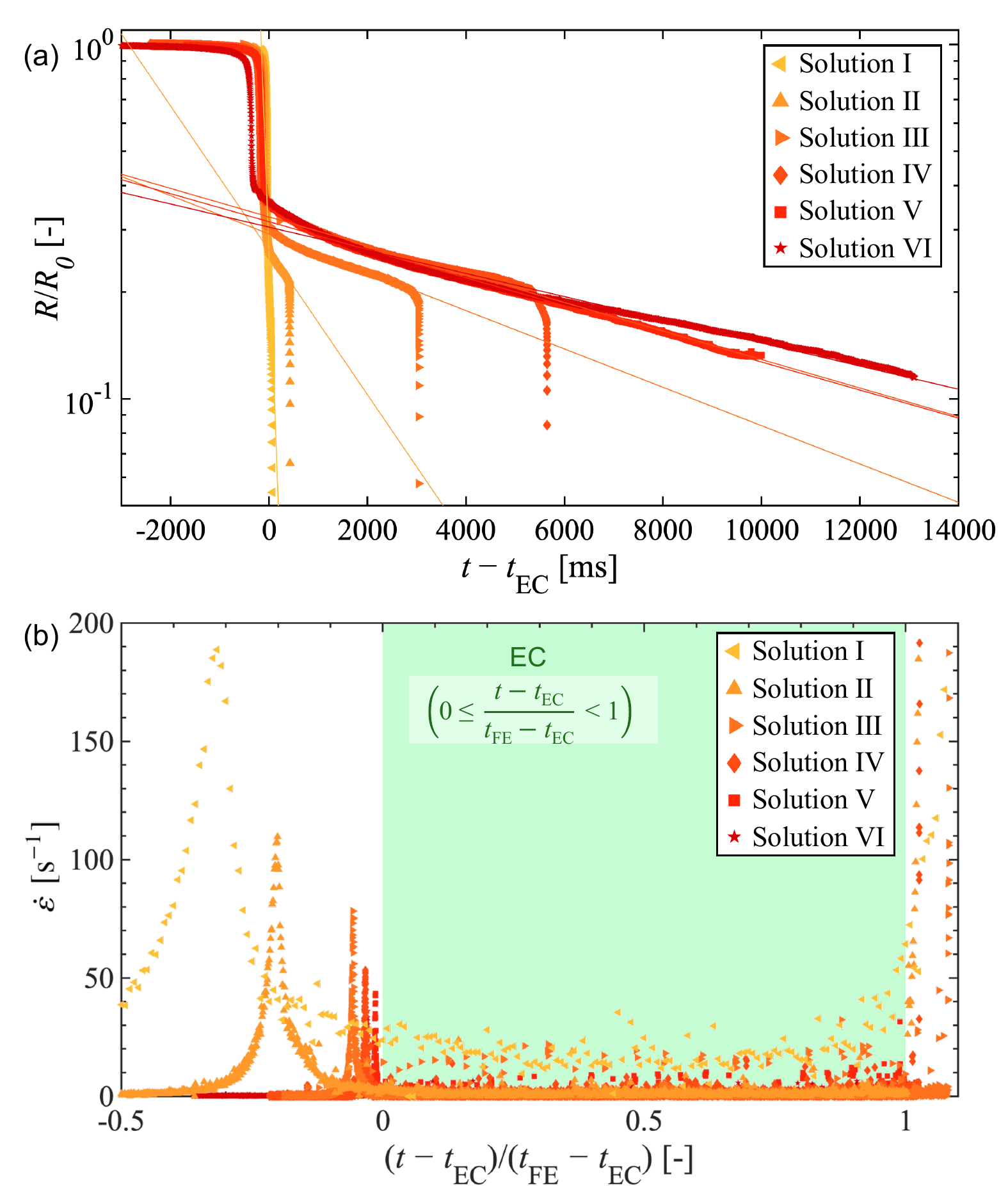}
\centering
\caption{Experimental results obtained from the light intensity images of filaments uniaxially extended by the liquid dripping method. Temporal evolution of (a) radius ratio $R/R_0$ and (b) extensional rate $\dot{\varepsilon}$ for micellar solutions $\mathrm I$ -- $\mathrm {V\hspace{-0.6pt}I}$.}
\label{radius_ratio}
\end{figure}

A semi-logarithmic graph of the filament radius ratio $R/R_0$ was obtained by analyzing the light intensity images (Fig.~\ref{intensity_retardation_images}(a)), is shown in Fig.~\ref{radius_ratio}(a). 
%===================================================
%===================================================
For all concentrations, after time $t - t_{\mathrm{EC}} = 0$ ms, the radius ratio transitioned into the EC regime, exhibiting an exponential decay over time. 
Moreover, IC/VC, EC, and TVEC regimes were confirmed for all solutions except $\mathrm {V}$ and $\mathrm {V\hspace{-0.6pt}I}$. 
For the solutions $\mathrm {V}$ and $\mathrm {V\hspace{-0.6pt}I}$, the TVEC regime was not observed because their magnification levels during the experiments were lower than those of the other solutions. 
This resulted in reduced spatial resolution (refer to Table \ref{Table:camera}), which was inadequate for capturing the breakage of the extending filaments within the TVEC regime.
In addition, Fig.~\ref{radius_ratio}(b) shows the plots of extensional rate $\dot{\varepsilon}$ versus nondimensional time, and it was confirmed that for all solutions, the extensional rate remained approximately constant within the EC regime compared to the IC/VC and TVEC regimes.
Here, the nondimensional time is defined as the normalized time $(t - t_{\mathrm{EC}})/(t_{\mathrm{FE}} - t_{\mathrm{EC}})$ [-], where $t_{\mathrm{FE}}-t_{\mathrm{EC}}$ denotes the extension durations, which represent the duration of the EC regime.
Table \ref{Table:C} lists the extension durations $t_{\mathrm{FE}}-t_{\mathrm{EC}}$ [ms] and extensional relaxation time $\lambda_{\mathrm{E}}$ [ms] for each solution. 
%===================================================
\begin{table}[tb!]
    \centering
    \caption{Extensional time range $t_\mathrm{\hspace{1pt} FE}-t_\mathrm{\hspace{1pt} EC}$, extensional relaxation time $\lambda_{\mathrm{E}}$, phase retardation at onset time of EC regime $\delta_\mathrm{\hspace{1pt} EC}$, and stress-optical coefficient $C$ under uniaxial extensional flow in CTAB/NaSal aqueous solutions with different micelle molar concentrations $c_\mathrm{V}$.}
    \label{Table:C}
    \tabcolsep = 0.2cm
    \renewcommand \arraystretch{1.35}
    \begin{tabular}{c c c c c c}
    \hline \hline
        Solution & $c_\mathrm{V}$ & $t_\mathrm{\hspace{1pt} FE}-t_\mathrm{\hspace{1pt} EC}$ & $\lambda_{\mathrm{E}}$ & $\delta_\mathrm{\hspace{1pt} EC}$ & $C$ $\times 10^{7}$ \\ 
        No. & [M] & [ms] & [ms] & [nm] & [Pa$^{-1}$] \\ \hline
        $\mathrm I$ & 0.005 & 60 & 379 & 3.9 & $-0.58 \pm 0.03$ \\ 
        $\mathrm {I\hspace{-0.6pt}I}$ & 0.010 & 410 & 667 & 10.9 & $-1.41 \pm 0.05$ \\ 
        $\mathrm {I\hspace{-0.6pt}I\hspace{-0.6pt}I}$ & 0.015 & 2800 & 2630 & 16.2 & $-2.04 \pm 0.13$ \\ 
        $\mathrm {I\hspace{-0.6pt}V}$ & 0.020 & 5500 & 3450 & 22.5 & $-2.98 \pm 0.17$ \\ 
        $\mathrm V$ & 0.025 & 9900 & 3520 & 25.7 & $-3.04 \pm 0.30$ \\  
        \hline
        $\mathrm {V\hspace{-0.6pt}I}$ & 0.030 & 12400 & 4120 & 28.7 & $-3.67 \pm 0.35$ \\ 
        \hline
        $\mathrm {V\hspace{-0.6pt}I\hspace{-0.6pt}I}$ & \multirow{11}{*}{0.030} & 68000 & 26700 & 32.1 & $-3.85 \pm 0.36$ \\ 
        $\mathrm {V\hspace{-0.6pt}I\hspace{-0.6pt}I \hspace{-0.6pt} I}$ & & 7000 & 12200 & 26.4 & $-3.63 \pm 0.06$ \\
        $\mathrm {I\hspace{-0.6pt}X}$ & & 6000 & 4130 & 30.2 & $-3.89 \pm 0.21$ \\
        $\mathrm {X}$ & & 14000 & 8060 & 30.0 & $-3.99 \pm 0.16$ \\
        $\mathrm {X\hspace{-0.6pt}I}$ & & 19200 & 6130 & 30.1 & $-3.79 \pm 0.38$ \\
        $\mathrm {X\hspace{-0.6pt}I\hspace{-0.6pt}I}$ & & 8700 & 4080 & 31.5 & $-3.78 \pm 0.37$ \\
        $\mathrm {X\hspace{-0.6pt}I\hspace{-0.6pt}I\hspace{-0.6pt}I}$ & & 10600 & 3180 & 31.6 & $-3.80 \pm 0.42$ \\
        $\mathrm {X\hspace{-0.6pt}I\hspace{-0.6pt}V}$ & & 3700 & 1650 & 28.5 & $-3.60 \pm 0.35$ \\
        $\mathrm {X\hspace{-0.6pt}V}$ & & 1500 & 581 & 27.8 & $-3.73 \pm 0.27$ \\
        $\mathrm {X\hspace{-0.6pt}V\hspace{-0.6pt}I}$ & & 780 & 303 & 28.3 & $-3.53 \pm 0.32$ \\
        $\mathrm {X\hspace{-0.6pt}V\hspace{-0.6pt}I\hspace{-0.6pt}I}$ & & 410 & 175 & 26.7 & $-3.53 \pm 0.11$ \\ \hline \hline
    \end{tabular}
\end{table}
%===================================================
Table \ref{Table:C} also includes the phase retardation $\delta_\mathrm{\hspace{1pt} EC}$ [nm] at the onset of the EC regime and the stress-optical coefficient $C$ [Pa$^{-1}$], as discussed in Section I\hspace{-1.2pt}I\hspace{-1.2pt}I. B. 
The extension duration increased with increasing micelle molar concentration $c_\mathrm{V}$ in the solution. 
An increase in the micelle molar concentration ($c_\mathrm{V}$) enhances the complex entanglements among micelles under extensional stress, thereby prolonging the extension duration ($t_{\mathrm{FE}} - t_{\mathrm{EC}}$) and extensional relaxation time ($\lambda_{\mathrm{E}}$).

The extensional stress $\sigma_{\rm \hspace{1pt} E}$ of each solution is calculated by substituting the measured minimum radius $R$ of the filament into Eq.~(\ref{stress in ECregime}). 
%===============================================
\begin{figure*}[t!]
\centering
\includegraphics[width=1\textwidth]{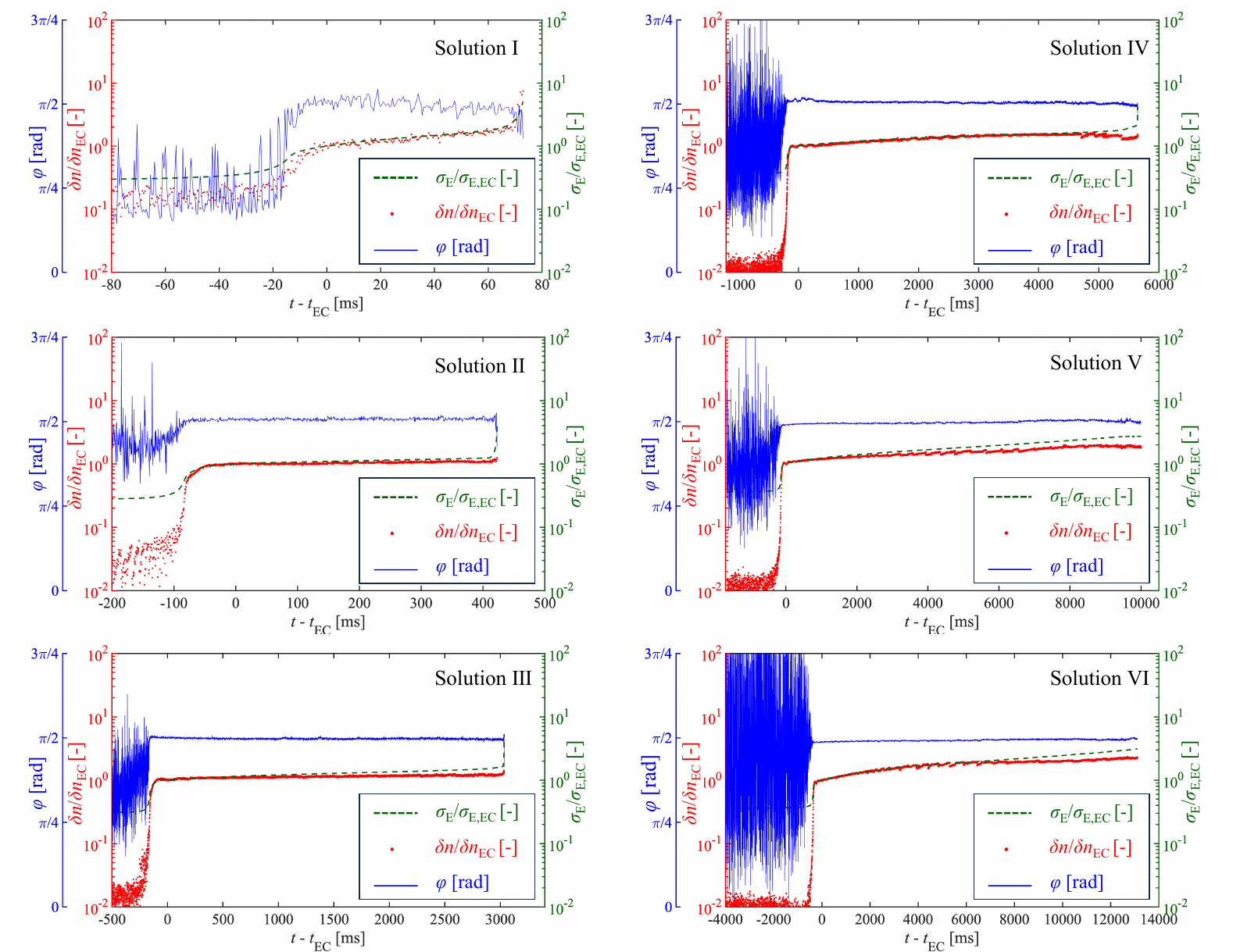}
\caption{Temporal evolution of normalized extensional stress $\sigma_{\rm \hspace{1pt} E}/\sigma_{\rm \hspace{1pt} E,EC}$, normalized birefringence $\delta n/\delta n_\mathrm{\hspace{1pt} EC}$, and orientation angle $\varphi$ for solutions $\mathrm I$ -- $\mathrm {V\hspace{-0.6pt}I}$.}
\label{Normalized_data}
\end{figure*}
%===============================================
Figure~\ref{Normalized_data} shows the normalized extensional stress $\sigma_{\rm \hspace{1pt} E}/\sigma_{\rm \hspace{1pt} E,EC}$ for all the solutions. 
Here, $\sigma_{\rm \hspace{1pt} E,EC}$ represents the extensional stress at the onset of the EC regime ($t - t_\mathrm{\hspace{1pt} EC} = 0$). 
In addition, the figure includes the normalized birefringence $\delta n/\delta n_\mathrm{\hspace{1pt} EC}$ and orientation angle $\varphi$, as discussed in Section I\hspace{-1.2pt}I\hspace{-1.2pt}I. B.
For all solutions, the normalized extensional stress increases in the IC/VC regime ($t - t_\mathrm{\hspace{1pt} EC} < 0$), remains relatively constant within the EC regime ($0 \leq t - t_\mathrm{\hspace{1pt} EC} < t_\mathrm{\hspace{1pt} FE} - t_\mathrm{\hspace{1pt} EC}$), and then sharply increases in the TVEC regime ($t - t_\mathrm{\hspace{1pt} EC} \geq t_\mathrm{\hspace{1pt} FE} - t_\mathrm{\hspace{1pt} EC}$). 
Theoretically, within the EC regime, the extensional stress continues to evolve, although its rate of change becomes significantly lower than that in the IC/VC and TVEC regimes.\cite{mckinley2005visco} 
This quasi-steady behavior allows for the reliable estimation of extensional stress from filament thinning dynamics, which is a key feature of elasto-capillary thinning.
Therefore, the present measurements of the extensional stress agree with this theory, confirming its validity.

\subsection{\label{birefringence result}Birefringence and orientation angle}
The measured birefringence and orientation angles were obtained by analyzing the measurement area within the phase retardation images, as shown in Fig.~\ref{intensity_retardation_images}(b). 
The measured phase retardation $\delta$ and orientation angles $\varphi$ were spatially averaged over the measurement area located at the central part of the filament where the radius was minimal, as indicated by the white square in Fig.~\ref{intensity_retardation_images}(b).
The size of this area ranged from $20 \times 20$ to $40 \times 40\ \mu\mathrm{m}^2$ (refer to Table \ref{Table:camera}), enabling high spatial resolution measurements.
The measured birefringence was obtained by dividing the measured phase retardation by the filament diameter ($\delta n = \delta/2R$).
Moreover, Fig.~\ref{intensity_retardation_images}(b) shows axial inhomogeneity in phase retardation along the filament, and such axial variation in extensional stress has also been reported in recent numerical simulations.\cite{zinelis2024fluid} 
Note that the phase retardation represents the integrated birefringence along the optical axis of the filament, and since birefringence is proportional to extensional stress (refer to Eqs.~(\ref{eq1}) and (\ref{eq2})), the location of maximum retardation does not necessarily coincide with the location of maximum extensional stress.
These findings emphasize that full-field 2D imaging is essential for capturing transient and localized stress-optical phenomena that may otherwise be missed due to spatial averaging.

It was expected that the lensing effect caused by the filament curvature would increase the light intensity, potentially leading to misinterpretation as apparent birefringence.
To minimize the influence of lensing effects, it is appropriate to spatiotemporally track the measurement area centered at the minimum radius, where the light path remains nearly straight.
To identify the appropriate measurement area, we analyzed the light intensity distribution near the filament center and selected the central area where the light intensity profile remained nearly uniform, even during the later stages of the EC regime when the filament curvature was at its maximum.
In the area, the transmitted light travels along straight paths without significant refraction, minimizing lensing effects and enabling the accurate application of the stress-optical rule under uniaxial extensional flow.
In addition, since the measurement area depends on the magnification used during imaging, the corresponding areas for each magnification, as listed in Table \ref{Table:camera}, were identified by examining light intensity images acquired at various magnifications for all solution conditions.
Note that we conducted an experiment using a non-birefringent fluid, a glycerol solution that lacks structural anisotropy such as rod-like or long-chain structures, and confirmed that no apparent birefringence increase was observed at the filament center due to the lensing effects.

The temporal evolution of the normalized birefringence $\delta n/\delta n_\mathrm{\hspace{1pt} EC}$ is shown in red in Fig.~\ref{Normalized_data}. 
Similar to the normalized extensional stress ($\sigma_{\rm E}/\sigma_{\rm E,EC}$), the measured birefringence ($\delta n$) was normalized by its value at the onset of the EC regime ($\delta n_\mathrm{\hspace{1pt} EC}$).
If the stress-optical coefficient remains constant, the normalized values of the extensional stress and birefringence should coincide, that is, $\sigma_{\rm \hspace{1pt} E}/\sigma_{\rm \hspace{1pt} E,EC} = \delta n/\delta n_\mathrm{\hspace{1pt} EC}$. 
Consequently, at all concentrations, the two quantities diverged in the IC/VC regime but began to increase just before the onset of the EC regime and tended to converge and coincide within it.  
This similarity in behavior suggests that within the EC regime, the stress-optical coefficient remains approximately constant. 
Although this does not confirm the strict validity of the stress-optical rule, it indicates that the stress-optical coefficient can be reasonably evaluated as a constant using the present technique.
Note that the birefringence measurements were also conducted for the CTAB aqueous solution ($c_\mathrm{D} = 0.10~\mathrm{M},\ c_\mathrm{S} = 0~\mathrm{M}$) and the NaSal aqueous solution ($c_\mathrm{D} = 0~\mathrm{M},\ c_\mathrm{S} = 0.360~\mathrm{M}$), and in both cases, no birefringence was observed.
Although the CTAB aqueous solution, exceeding the critical micelle concentration, likely contains spherical micelles, the absence of birefringence suggests that optical anisotropy requires the presence of high aspect ratio structures such as wormlike micelles.

Orientation angle $\varphi$ (blue line in Fig.~\ref{Normalized_data}) is based on a horizontal reference of $0$ rad, with the micelles oriented in the direction of extension at $\pi/2$ rad.
At the initial stage of extension in the IC/VC regime ($t - t_\mathrm{\hspace{1pt} EC} < 0$), $\varphi$ oscillates temporally and is unstable across all concentrations, indicating a statistically random orientation of the micelles.
However, immediately before entering the EC regime, $\varphi$ begins converging, whereas within the EC regime ($0 \leq t - t_\mathrm{\hspace{1pt} EC} < t_\mathrm{\hspace{1pt} FE} - t_\mathrm{\hspace{1pt} EC}$), it settles at a constant $\pi/2$ rad. 
Therefore, within the EC regime, where the orientation angle remains constant at $\pi/2$ rad, the micelles are statistically highly oriented in the extension direction.
Interestingly, the convergence of the orientation angle towards a constant value ($\pi/2$ rad) begins even before the EC regime, indicating that the orientation of the micelles towards the extension direction begins earlier than in the uniaxially extended state within the EC regime. 
In the TVEC regime ($t - t_\mathrm{\hspace{1pt} EC} \geq t_\mathrm{\hspace{1pt} FE} - t_\mathrm{\hspace{1pt} EC}$), where the filament extends further until breaking, the orientation angle becomes unstable because the effect of light refraction owing to the curvature of the liquid filament becomes more pronounced.
Although the birefringence and orientation angle data suggest a high degree of alignment along the extensional direction, the micelles were not uniformly oriented; rather, they exhibited a statistically high orientation.
It is possible that the micelles followed a Gaussian random distribution biased toward the extensional direction.
However, the degree of this statistical alignment could not be quantitatively evaluated in the present study and is left for future investigation.

The orientation of micelles along the filament extension resulted in two key findings.  
First, the birefringence observed under these conditions can be interpreted as the intrinsic birefringence.  
The flow birefringence observed in complex fluids under deformation results from the optical anisotropy induced by the alignment of microstructural units, such as polymers or micelles.  
This macroscopic optical anisotropy can be interpreted as the projection of the intrinsic birefringence of individual molecules or assemblies along the observation direction, weighted by their orientation distribution (e.g., derived from the orientation angle).\cite{Larson1999}  
Intrinsic birefringence refers to the molecular-level optical anisotropy inherent to an entity, such as a wormlike micelle, owing to its anisotropic polarizability along its long axis.  
Conversely, flow birefringence is a macroscopic manifestation of the intrinsic birefringence of oriented microscopic structures.  
The relationship between the flow birefringence $\delta n$ and orientation angle $\varphi$ is expressed as follows:\cite{okada2016reliability}
\begin{equation}
\delta n = f \delta n^{0} = \frac{3\cos^{2}(\pi/2-\varphi)-1}{2} \delta n^{0}
\label{orientation_angle},
\end{equation}
where $\delta n^{0}$ [–] represents the intrinsic birefringence, which is a material constant for the birefringence of fully oriented micelles, and $f$ [–] is alignment parameter.
Importantly, the intrinsic birefringence ($f \sim 1$) is considered within the EC regime because the orientation angle remains constant at $\pi/2$.

Second, because the principal direction of the extensional stress ($\sigma_{\perp}$) coincides with the direction of the major axis of the refractive index ($n_{\perp}$), the stress-optical relationship (Eq.~(\ref{eq1})) yields $\delta n = n_{\parallel} - n_{\perp} < 0$, indicating negative birefringence.  
Accordingly, the stress-optical coefficient ($C$) for CTAB/NaSal solutions was consistently negative across all conditions tested.  
This result coincides exactly with previous findings for worm-like micelles under shear flow, where similar negative values of $C$ were reported because of the alignment of micelles along the stress direction.\cite{shikata1994rheo,wheeler1996structure,C_Takahashi}  

In Fig.~\ref{Normalized_data}, the orientation angle for solution $\mathrm I$ ($c_\mathrm{D}=0.005$ M, $c_\mathrm{S}=0.038$ M) does not fully converge to $\pi/2$ rad and shows significant fluctuations within the EC regime.
This is because solution $\mathrm I$, which has a lower micelle molar concentration ($c_\mathrm{V}$) than the other solutions, makes the measurement challenging.
The results confirmed that the solution $\mathrm I$ exhibited a small phase retardation of $\delta _\mathrm{\hspace{1pt} EC}=3.9$ nm (Table \ref{Table:C}). 
Considering that the measurement limit for phase retardation using a high-speed polarization camera is approximately 5 nm, the camera cannot accurately capture the minor phase retardation induced by birefringence in the solution $\mathrm I$.

\subsection{\label{birefringence}Stress-optical coefficient}
\subsubsection{Validity of resulting stress-optical coefficient}
This section discusses the derivation and validity of the stress-optical coefficient based on the measured extensional stress and birefringence.
To demonstrate the validity of the stress-optical rule under shear flow, Takahashi et al. \cite{C_Takahashi} reported the need to verify two points: (A) the proportionality between the refractive index and stress tensors ($\bm{n} = C \bm{\sigma}$), and (B) the constancy of the stress-optical coefficient regardless of the shear rate. 
Here, we examine these two points using experimental data obtained under uniaxial extensional stress conditions, and evaluate the validity of the resulting stress-optical coefficient.

\begin{figure}[tb!]
\centering
\includegraphics[width=\columnwidth]{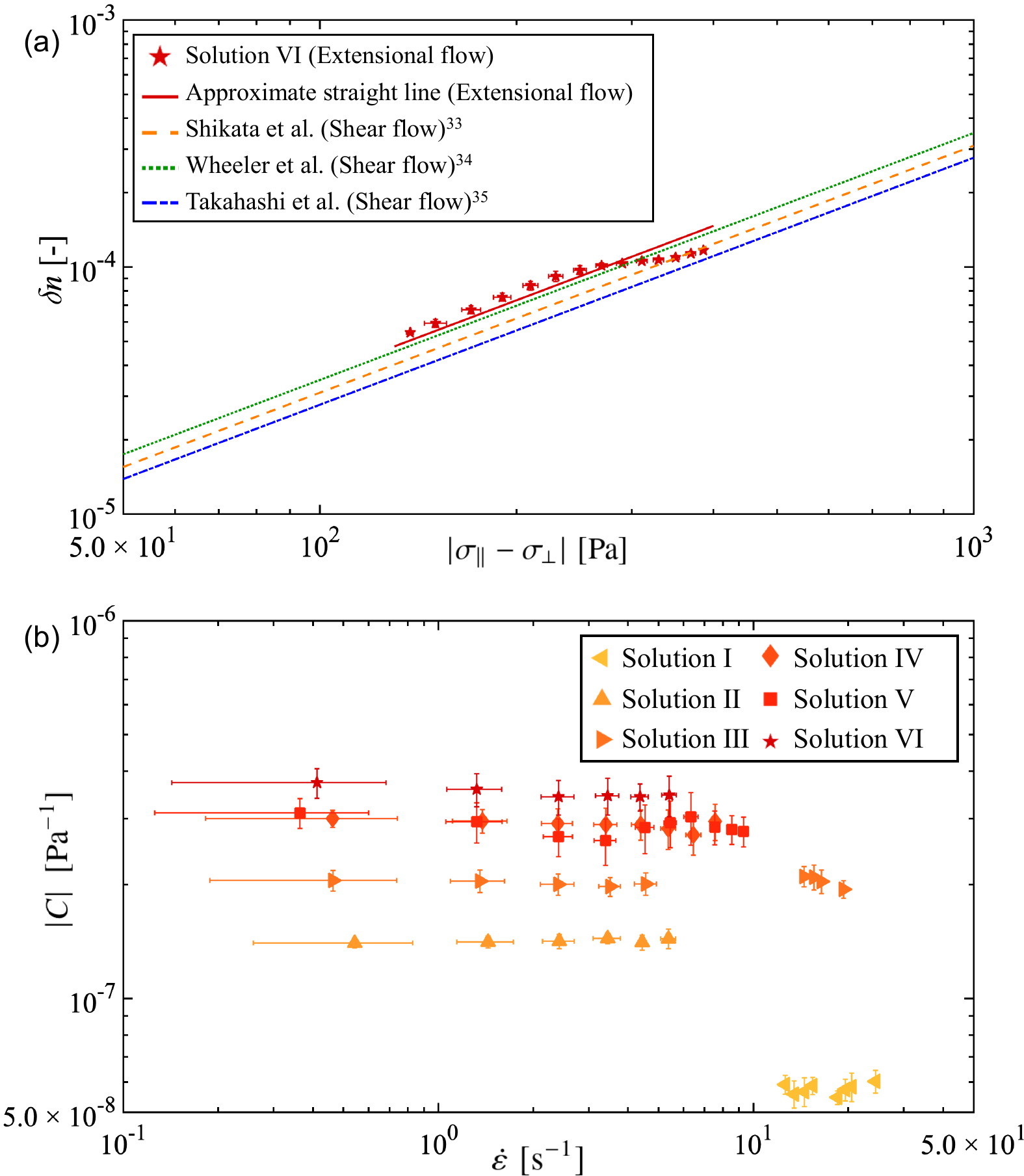}
\caption{Experimental evidence for the validity of the stress-optical rule under uniaxial extensional stress condition. 
(a) birefringence $\delta n$ vs. principal stress difference $|\sigma_{\parallel}-\sigma_{\perp}|$ for solution $\mathrm {V\hspace{-0.6pt}I}$ within EC regime. The plots and error bars indicate the mean $\delta n$ and standard deviation of each plot within a bandwidth of 20 $\rm Pa$ for the extensional stress, respectively. 
(b) stress-optical coefficient $|C|$ vs. extensional rate $\dot{\varepsilon}$ within EC regime. The plots and error bars indicate the mean $|C|$ and standard deviation of each plot within a bandwidth of 1 s$^{-1}$ for the extension rate, respectively.}
\label{C}
\end{figure}

To investigate point (A), the linear relationship between birefringence $\delta n$ and the principal stress difference $|\sigma_{\parallel}-\sigma_{\perp}|$ was examined using the data of $\delta n$ and the extensional stress $\sigma_{\rm \hspace{1pt} E}$ within the EC regime obtained by the present technique.
To enable comparison with previous studies,\cite{shikata1994rheo,wheeler1996structure,C_Takahashi} Fig.~\ref{C}(a) presents the graph showing the relationship between birefringence and extensional stress for solution $\mathrm {V\hspace{-0.6pt}I}$ ($c_\mathrm{D}=0.030$ M and $c_\mathrm{S}=0.230$ M).
%===============================================
%===============================================
The result confirmed that the birefringence increased monotonically with extensional stress, indicating a linear relationship between the two; although both quantities slightly increase within the EC regime, their proportional relationship is maintained, leading us to conclude that strain-hardening does not occur under these conditions.
Moreover, the slope obtained from the measurement data for the stress-optical coefficient $|C|$ was found to be consistent in order of magnitude with the values reported in previous studies.\cite{shikata1994rheo,wheeler1996structure,C_Takahashi}
In the present study, the measured stress-optical coefficient under uniaxial extensional flow was $C=-3.67 \times 10^{-7}$ Pa$^{-1}$. 
This value was closest to that reported by Wheeler et al.\cite{wheeler1996structure} ($C=-3.49 \times 10^{-7}$ Pa$^{-1}$) and approximately agreed with the values reported by Shikata et al.\cite{shikata1994rheo} and Takahashi et al.\cite{C_Takahashi} under shear flow ($C=-3.1 \times 10^{-7}$ Pa$^{-1}$ and $C=-2.77 \times 10^{-7}$ Pa$^{-1}$, respectively). 
Therefore, the stress-optical coefficients of micellar solutions were consistent under shear and extensional flows.
The means and standard deviations of the stress-optical coefficients in the other solutions are listed in Table \ref{Table:C}. 
The small standard deviations indicate the high reliability of the present technique.

To investigate Point (B), the stress-optical coefficients and extensional rates were compared for each solution.
Note that the present study employed the "extension rate" instead of the "shear rate" used in previous studies.\cite{shikata1994rheo,wheeler1996structure,C_Takahashi}
Because both the shear and extension rates are components of the strain-rate tensor and represent the deformation-induced orientation of microstructures, the extension rate was considered equally valid for assessing the stress-optical coefficient.  
Therefore, the approach for evaluating the constancy of the stress-optical coefficient with respect to the extension rate was physically justified.  
Figure~\ref{C}(b) shows the stress-optical coefficient $|C|$ [Pa$^{-1}$] derived from $|\delta n / \sigma_{\rm \hspace{1pt} E}|$ within the EC regime ($0 \leq t - t_\mathrm{\hspace{1pt} EC} < t_\mathrm{\hspace{1pt} FE} - t_\mathrm{\hspace{1pt} EC}$) for each solution.
The horizontal axis represents the extension rate $\dot{\varepsilon}$ [s$^{-1}$] within the EC regime for each solution. 
The stress-optical coefficient $|C|$ was confirmed to remain constant regardless of the extension rate $\dot{\varepsilon}$, indicating that it can be appropriately defined as an intrinsic material property.

These results confirm that the measurements obtained using the present technique fully satisfy points (A) and (B) for validating the stress-optical rule identified by Takahashi et al. \cite{C_Takahashi} 
The above discussion demonstrates that the stress-optical coefficients obtained for micellar solutions under uniaxial extensional flows are consistent with those reported for shear flows.\cite{shikata1994rheo,wheeler1996structure,C_Takahashi} 
Note that we successfully measured the stress-optical coefficient without observing any breakdown in the stress-optical rule, even at strain rates significantly higher than those explored in previous studies.\cite{shikata1994rheo,wheeler1996structure,C_Takahashi}
For reference, the stress-optical coefficients in previous reports were obtained under shear flow at relatively low shear rates: $0.16 < \dot{\gamma} < 2.5~\mathrm{s^{-1}}$ in Shikata et al.,\cite{shikata1994rheo} $0.15 < \dot{\gamma} < 3.0~\mathrm{s^{-1}}$ in Wheeler et al.,\cite{wheeler1996structure} 
and $0.36 < \dot{\gamma} < 3.6~\mathrm{s^{-1}}$ in Takahashi et al.\cite{C_Takahashi}
Moreover, Takahashi et al.\cite{C_Takahashi} reported deviations from the stress-optical rule under shear flow at shear rates exceeding $\dot{\gamma} > 4~\mathrm{s^{-1}}$. 
In contrast, the present study demonstrated that reliable stress-optical coefficients can be obtained over a broad range of extension rates ($0.1 < \dot{\varepsilon} < 30~\mathrm{s^{-1}}$), indicating that the stress-optical rule remains valid even at relatively high strain rates under extensional flow conditions.
This difference is attributed to the inherent advantages of uniaxial extensional flow, as described in Section I\hspace{-1.2pt}I. C. 
In such extensional flows, strong molecular alignment along the extension direction results in enhanced anisotropy in the optical index ellipsoid, thereby enabling more precise optical measurements compared to shear flow. 

Here, we compare the present study with the influential work of Rothstein,~\cite{rothstein2003transient} who made significant contributions to this field.
He reported, through careful experimentation, that CTAB/NaSal solutions may not strictly follow the stress-optical rule under uniaxial extensional flow at relatively high extension rates of $0.5 < \dot{\varepsilon} < 4.5~\mathrm{s^{-1}}$.
First, we would like to emphasize that the CTAB/NaSal micellar solutions used in the present study differ from those used by Rothstein in terms of molar concentration and molar ratio. 
He focused on the solutions with the CTAB/NaSal molar ratio of $0.5 \leq c_\mathrm{S}/c_\mathrm{D} \leq 2$, which is known to produce worm-like micellar structures and investigated strain-hardening behavior under those conditions.
In contrast, the present study covers a broader range of molar ratio ($1 \leq c_\mathrm{S}/c_\mathrm{D} \leq 6$), which includes not only wormlike micelles but also particularly the regime where network-like structures form. 
These structural differences in the micellar networks may influence the degree of molecular orientation and, consequently, the applicability of the stress-optical rule.

\begin{figure}[tb!]
\centering
\includegraphics[width=\columnwidth]{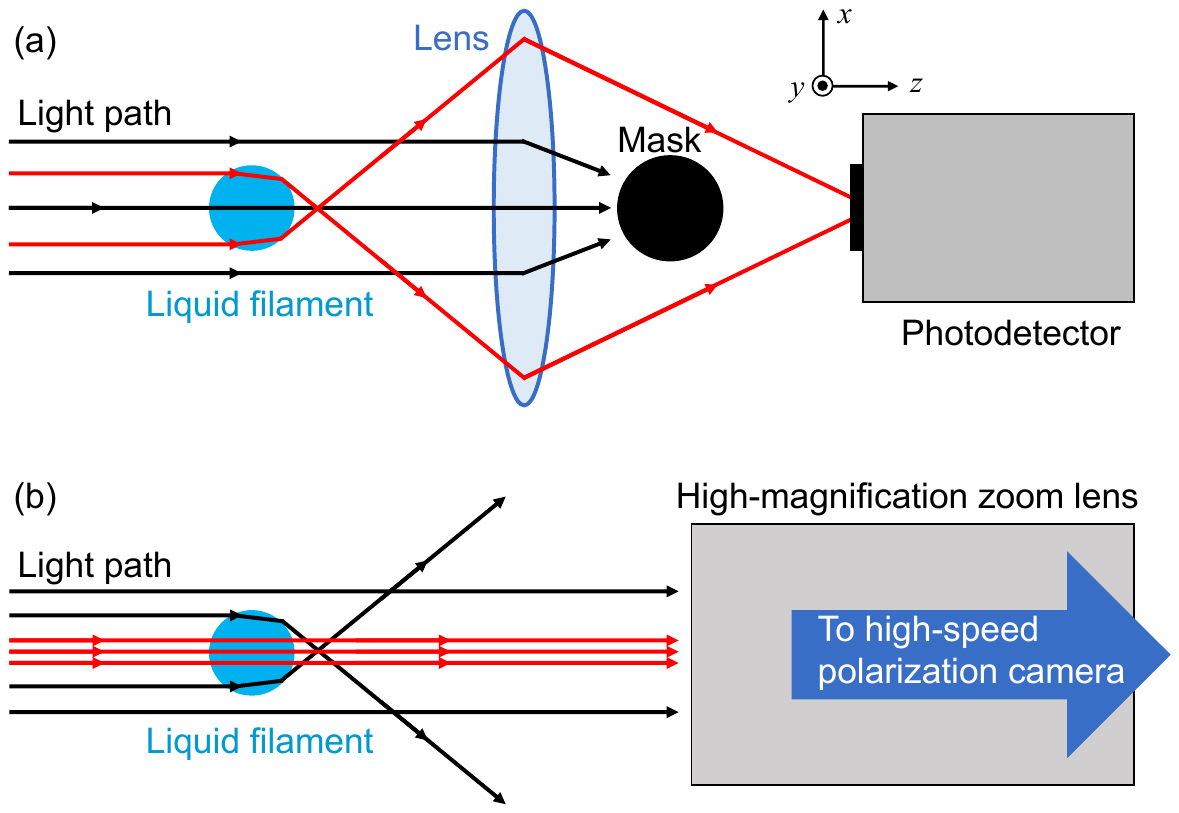}
\caption{Optical path comparison between (a) a conventional pointwise measurement method using a laser and photodetector as reported by Rothstein et al.,\cite{rothstein2003transient,rothstein2002comparison, rothstein2002inhomogeneous} and (b) the present full-field technique using a high-speed polarization camera. In the pointwise method, light rays refracted by the filament (indicated by the red line) are collected by a convex lens and directed to the photodetector. On the other hand, the present full-field technique extracts only the light rays that travel straight through the center of the filament (indicated by the red line) using image analysis.}
\label{optics}
\end{figure}

Moreover, the experimental measurement technique in the previous study reported by Rothstein et al.\cite{rothstein2003transient,rothstein2002comparison, rothstein2002inhomogeneous} differed from those of the present study, particularly in the use of the conventional pointwise measurements employing laser probing, as shown in Fig.~\ref{optics}(a). 
%===============================================
%===============================================
In the conventional pointwise measurements, the birefringence was measured using a laser-based setup in which only the refracted light rays through the filament was detected by a photodetector via a lens and a mask. 
However, this approach spatially averages the signal of the refracted light rays and may exclude the straight central light rays that contain the most accurate birefringence information, potentially reducing measurement accuracy.
On the other hand, our technique combines full-field 2D imaging with post-processing, allowing for selective extraction of the straight central light rays least affected by the lensing effects, as shown in Fig.~\ref{optics}(b). 
This enables more accurate determination of both birefringence and orientation angle.

Such conventional pointwise measurements also had limitations in capturing localized birefringence in the region of maximum extensional stress, which typically occurs near the filament’s minimum radius and gradually shifts its position over time.
In the previous study reported by Rothstein,\cite{rothstein2003transient} the temporal and spatial resolutions of the pointwise measurements were limited to 40 fps and 20 $\mu$m, respectively, which may have introduced a noticeable time lag between the development of extensional stress and the birefringence response.
These limitations in spatiotemporal resolution could have contributed to the apparent deviations from the stress-optical rule observed in his study. 
On the other hand, the present technique, which employs the high-speed polarization camera capable of recording full-field 2D birefringence images at up to 2000 fps with a maximum spatial resolution of 5 $\mu$m, offers enhanced temporal and spatial resolution for more detailed analysis of the correspondence between birefringence and extensional stress.
Note that we do not claim the universal validity of the stress-optical rule in extensional flow based on the present experimental data.
Rather, our findings demonstrate that under the specific experimental conditions, i.e., the tested CTAB/NaSal solutions and extension rate range, the stress-optical rule rule remained valid and that this validity was confirmed through high-resolution measurements.

A limitation of the present technique is that rheo-optical measurements become invalid once the liquid filament stretches beyond the field of view of the high-speed polarization camera. 
Since the position of the minimum radius, where extensional stress is concentrated, must be captured to maintain the stress–birefringence correspondence, failure to image this area compromises the analysis.
Therefore, we conclude that the discrepancies between the present study and previous study \cite{shikata1994rheo,wheeler1996structure,C_Takahashi,rothstein2003transient} are likely attributable to differences in molecular orientation dynamics and the spatiotemporal resolution of the respective measurement techniques. 
For this reason, direct comparisons should be made with appropriate consideration of these differences.

\subsubsection{Micelle concentration dependence of the stress-optical coefficient}
\begin{figure}[tb!]
\centering
\includegraphics[width=\columnwidth]{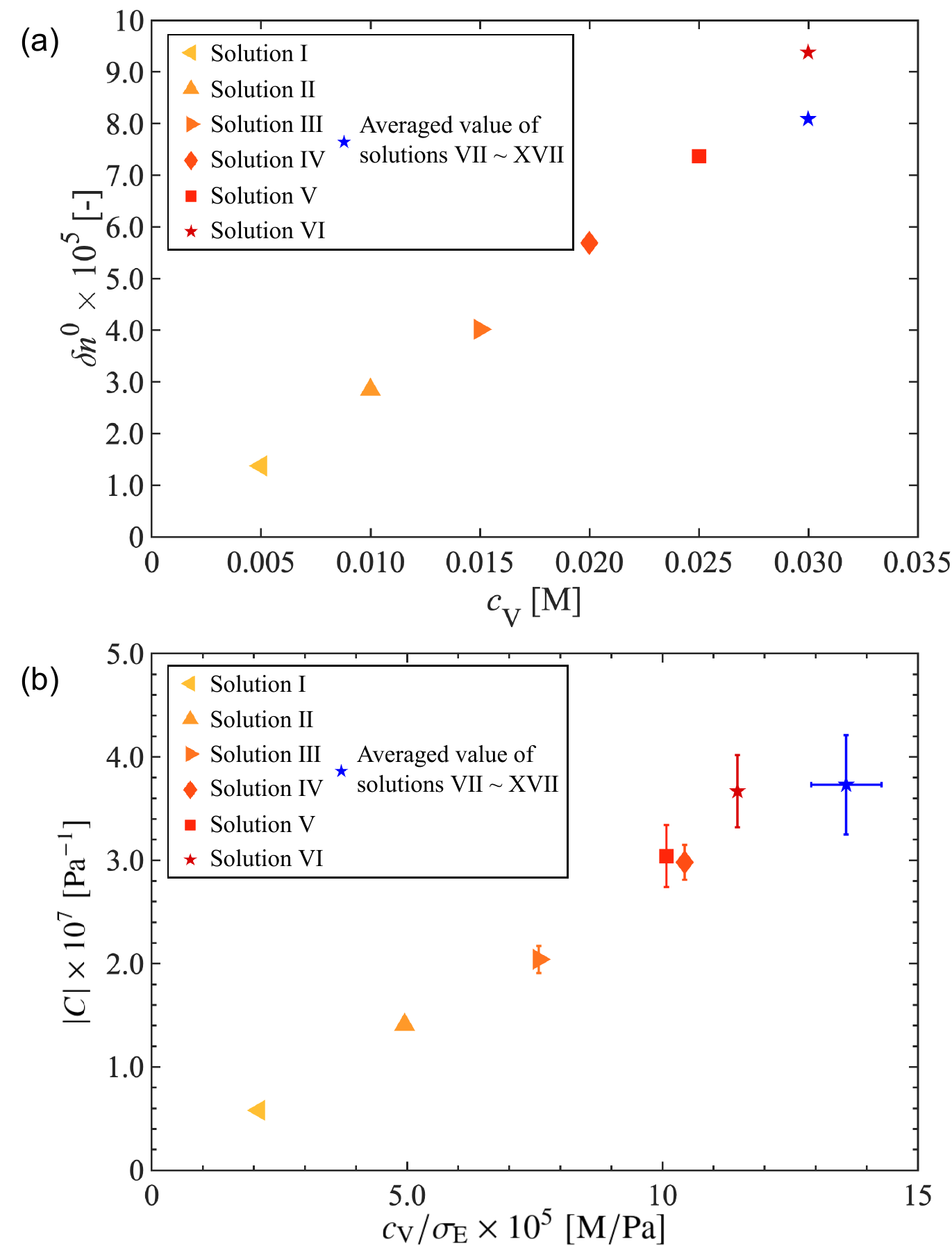}
\caption{(a) Dependence of the intrinsic birefringence $\delta n^{0}$ on the micelle molar concentration $c_\mathrm{V}$, and (b) Dependence of the stress-optical coefficient $|C|$ on the ratio of the micelle molar concentration to the extensional stress $c_\mathrm{V}/\sigma_{\rm E}$.}
\label{cV_C}
\end{figure}

To further investigate the reliability of the obtained stress-optical coefficient for uniaxially extending micellar solutions, we consider the factors influencing the stress-optical coefficient.  
Assuming that the micelles behave as optically anisotropic rigid rods, the intrinsic birefringence $\delta n^{0}$ [–] can be expressed as follows:\cite{cerf1952flow,wunderlich1987flow,shikata1994rheo,DoiEdwards1986,fuller1995optical}
\begin{equation}
\delta n^{0} = \frac{2 \pi(\overline{n}^2+2)^2 N_\mathrm{A} c_\mathrm{V} \Delta \alpha^0}{9\overline{n} m},
\label{eq:birefringence_model}
\end{equation}
where $\overline{n}$ [–] is the average refractive index of the solution, $N_\mathrm{A}$ [mol$^{-1}$] is Avogadro’s number ($N_\mathrm{A} = 6.02 \times 10^{23}$ mol$^{-1}$), and $\Delta \alpha^0$ [$\text{\AA}^3$] is the intrinsic polarizability anisotropy of a monomer unit (i.e., Kuhn segment) in a flexible micellar chain.
In this model, the wormlike micelle is constructed by disk-shaped monomers consisting of $m$ CTA$^+$Sal$^-$ complexes.\cite{shikata1994rheo}
In the present study, these parameters ($\overline{n}$, $N_\mathrm{A}$, $\Delta \alpha^0$, and $m$) are assumed to be constant.  
As indicated by Eq.~(\ref{eq:birefringence_model}), the intrinsic birefringence $\delta n^{0}$ is proportional to the micelle molar concentration $c_\mathrm{V}$, and thus to the number of oriented micelles per unit volume, given by $N_\mathrm{A} c_\mathrm{V} / m$. 
Indeed, as shown in Fig.~\ref{cV_C}(a), a linear relationship was observed between $\delta n^{0}$ and $c_\mathrm{V}$, which is consistent with the experimental trend under shear flow reported by Shikata et al.\cite{shikata1994rheo}
Here, since the alignment parameter was $f \sim 1$ in uniaxial extensional flows for all solutions, the measured birefringence can be regarded as the intrinsic birefringence. 
Accordingly, the intrinsic birefringence value was defined as the average birefringence measured within the EC regime.
%===================================================
%===================================================

Moreover, Shikata et al.\cite{shikata1994rheo} reported the intrinsic polarizability anisotropy of wormlike micelles to be approximately $|\Delta \alpha^0| \sim 14~\text{\AA}^3$ for the solution with $c_\mathrm{V} = 0.030$ M ($c_\mathrm{D} = 0.030$ M, $c_\mathrm{S} = 0.230$ M) and $m \sim 18$.
Using our experimental data for the same solution (solution $\mathrm{V\hspace{-0.6pt}I}$), we also estimated the intrinsic polarizability anisotropy based on the measured values of $\overline{n} = 1.342$ and $\delta n^{0} = 9.38 \times 10^{-5}$, obtained using a standard Abbe refractometer (ER-2S, Matsuyoshi Medical Instruments Co.) and the high-speed polarization camera, respectively.
As a result, we obtained $|\Delta \alpha^0| \sim 12~\text{\AA}^3$, which is in good agreement with the literature value.\cite{shikata1994rheo}

By combining Eqs.~(\ref{orientation_angle}) and (\ref{eq:birefringence_model}) with the definition of the stress-optical coefficient (Eq.~(\ref{eq4})), the following expression is obtained:
\begin{equation}
C = f \frac{2 \pi(\overline{n}^2+2)^2 N_\mathrm{A} c_\mathrm{V} \Delta \alpha^0}{9\overline{n} m \sigma_{\rm E}}.
\label{eq:stress_optical_coefficient_model}
\end{equation}
Since the intrinsic birefringence ($\delta n^{0}$ at $f \sim 1$) is proportional to the micelle molar concentration, Eq.~(\ref{eq:stress_optical_coefficient_model}) indicates that the stress-optical coefficient $C$ is proportional to the ratio of the micelle molar concentration to the extensional stress $c_\mathrm{V}/\sigma_{\rm E}$.  
As shown in Fig.~\ref{cV_C}(b), the absolute value of the stress-optical coefficient $|C|$ increases monotonically with increasing $c_\mathrm{V}/\sigma_{\rm E}$.  
Despite the simplifying assumptions, the observed linear relationship in Figs.~\ref{cV_C}(a) and \ref{cV_C}(b) provides strong support for the validity of the present technique.

\begin{figure}[tb!]
\centering
\includegraphics[width=0.96\columnwidth]{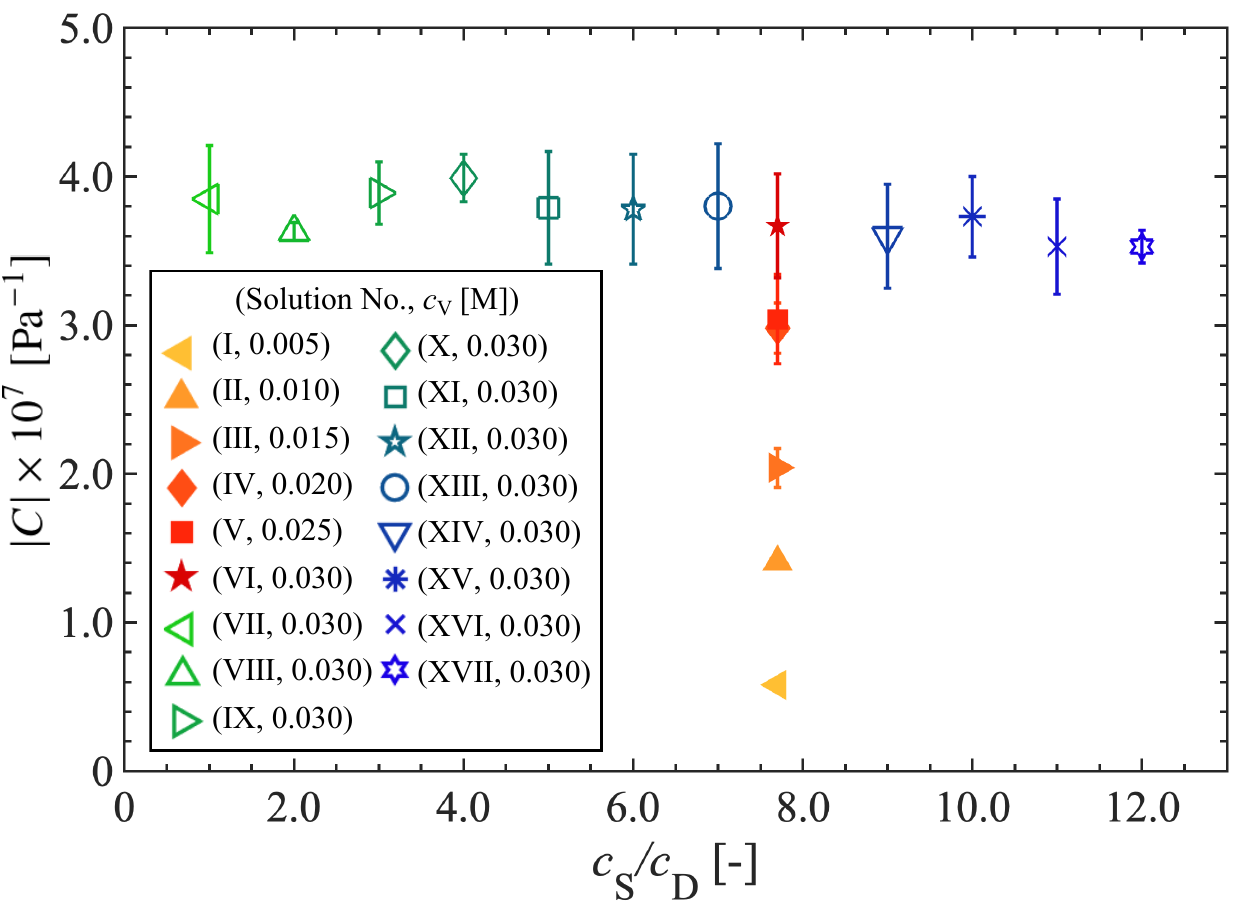}
\caption{Dependence of stress-optical coefficient $|C|$ on CTAB/NaSal molar ratio $c_\mathrm{S} / c_\mathrm{D}$.}
\label{cS/cD_C}
\end{figure}

Furthermore, the dependence of the CTAB/NaSal molar ratio $c_\mathrm{S} / c_\mathrm{D}$ on the stress-optical coefficient $C$ was discussed.
As shown in Fig.~\ref{cS/cD_C}, the warm-colored plots represent the results for solutions $\mathrm I$ to $\mathrm {V\hspace{-0.6pt}I}$, in which the micelle molar concentration was varied at a constant CTAB/NaSal molar ratio, while the cool-colored plots represent solutions $\mathrm{V\hspace{-0.6pt}I \hspace{-0.6pt} I}$ to $\mathrm {X\hspace{-0.6pt}V\hspace{-0.6pt}I \hspace{-0.6pt} I}$, in which the CTAB/NaSal molar ratio was varied at a constant volume fraction.
%===============================================
%===============================================
As shown in the figure, the stress-optical coefficient depends on the micelle molar concentration $c_\mathrm{V}$ and is almost independent of the CTAB/NaSal molar ratio $c_\mathrm{S} / c_\mathrm{D}$.
An increase in micelle molar concentration indicates a greater number of oriented micelles, which consequently  results in a higher stress-optical coefficient.
However, even if the CTAB/NaSal molar ratio alters the micelle length or cross-linked structures, the stress-optical coefficient remains unchanged, provided the intrinsic optical anisotropy of each micelle and its orientational response to external forces remain essentially the same.
This result strongly suggests that the stress-optical response is governed primarily by the number of oriented micelles and their average degree of orientation rather than by micellar morphology.
This provides physical support for the view that the birefringence is predominantly determined by the formation of a macroscopic orientational field rather than by subtle variations in micellar structure.

\section{Conclusions}
The unsteady structural behavior of CTAB/NaSal micellar solutions with varying micellar morphologies under uniaxial extensional stress was investigated using a newly developed "full-field extensional rheo-optical technique," which integrates the liquid dripping method and a high-speed polarization camera. 
This approach enabled the simultaneous and noncontact measurement of the extensional stress and full-field 2D flow birefringence with high spatiotemporal resolution.

The observed birefringence and orientation angle data revealed that micelles became increasingly oriented along the extensional direction as the deformation proceeded. 
Within the EC regime, the birefringence and extensional stress evolved consistently, and their linear correlation suggested that the stress-optical coefficient remained approximately constant during this regime. 
The orientation angle also converged to a stable value aligned with the extensional axis, indicating a high degree of statistical alignment and supporting the interpretation that the measured birefringence reflects the intrinsic birefringence.

Comparative analysis across multiple concentrations showed that the stress-optical coefficient increased linearly with the ratio of micelle molar concentration to extensional stress, indicating that the birefringence response was governed by the number of oriented micelles. 
In contrast, the CTAB/NaSal molar ratio had a negligible effect, implying that changes in micellar morphology did not significantly influence the optical response in the current deformation regime.
Thus, the present technique provides a robust and quantitative framework for probing anisotropic molecular responses in complex fluids under extensional flows, with potential applications in both soft matter physics and material design.

\begin{acknowledgments}
The present study was supported by the Japan Society for the Promotion of Science, KAKENHI Grant Nos. 23H01343 and was partially based on the results obtained from project JPNP20004, subsidized by the New Energy and Industrial Technology Development Organization (NEDO).
\end{acknowledgments}

\section*{Data Availability Statement}
The data that support the findings of the present study are available from the corresponding author upon reasonable request.

\section*{References}
%\nocite{*}
\bibliography{aipsamp}% Produces the bibliography via BibTeX.

\providecommand{\noopsort}[1]{}\providecommand{\singleletter}[1]{#1}%
\begin{thebibliography}{81}%
\makeatletter
\providecommand \@ifxundefined [1]{%
 \@ifx{#1\undefined}
}%
\providecommand \@ifnum [1]{%
 \ifnum #1\expandafter \@firstoftwo
 \else \expandafter \@secondoftwo
 \fi
}%
\providecommand \@ifx [1]{%
 \ifx #1\expandafter \@firstoftwo
 \else \expandafter \@secondoftwo
 \fi
}%
\providecommand \natexlab [1]{#1}%
\providecommand \enquote  [1]{``#1''}%
\providecommand \bibnamefont  [1]{#1}%
\providecommand \bibfnamefont [1]{#1}%
\providecommand \citenamefont [1]{#1}%
\providecommand \href@noop [0]{\@secondoftwo}%
\providecommand \href [0]{\begingroup \@sanitize@url \@href}%
\providecommand \@href[1]{\@@startlink{#1}\@@href}%
\providecommand \@@href[1]{\endgroup#1\@@endlink}%
\providecommand \@sanitize@url [0]{\catcode `\\12\catcode `\$12\catcode `\&12\catcode `\#12\catcode `\^12\catcode `\_12\catcode `\%12\relax}%
\providecommand \@@startlink[1]{}%
\providecommand \@@endlink[0]{}%
\providecommand \url  [0]{\begingroup\@sanitize@url \@url }%
\providecommand \@url [1]{\endgroup\@href {#1}{\urlprefix }}%
\providecommand \urlprefix  [0]{URL }%
\providecommand \Eprint [0]{\href }%
\providecommand \doibase [0]{http://dx.doi.org/}%
\providecommand \selectlanguage [0]{\@gobble}%
\providecommand \bibinfo  [0]{\@secondoftwo}%
\providecommand \bibfield  [0]{\@secondoftwo}%
\providecommand \translation [1]{[#1]}%
\providecommand \BibitemOpen [0]{}%
\providecommand \bibitemStop [0]{}%
\providecommand \bibitemNoStop [0]{.\EOS\space}%
\providecommand \EOS [0]{\spacefactor3000\relax}%
\providecommand \BibitemShut  [1]{\csname bibitem#1\endcsname}%
\let\auto@bib@innerbib\@empty
%</preamble>
\bibitem [{\citenamefont {Philippoff}, \citenamefont {Gaskins},\ and\ \citenamefont {Brodnyan}(1957)}]{philippoff1957flow}%
  \BibitemOpen
  \bibfield  {author} {\bibinfo {author} {\bibfnamefont {W.}~\bibnamefont {Philippoff}}, \bibinfo {author} {\bibfnamefont {F.~H.}\ \bibnamefont {Gaskins}}, \ and\ \bibinfo {author} {\bibfnamefont {J.~G.}\ \bibnamefont {Brodnyan}},\ }\bibfield  {title} {\enquote {\bibinfo {title} {Flow birefringence and stress. {V}. {C}orrelation of recoverable shear strains with other rheological properties of polymer solutions},}\ }\href@noop {} {\bibfield  {journal} {\bibinfo  {journal} {Journal of Applied Physics}\ }\textbf {\bibinfo {volume} {28}},\ \bibinfo {pages} {1118--1123} (\bibinfo {year} {1957})}\BibitemShut {NoStop}%
\bibitem [{\citenamefont {Chow}\ and\ \citenamefont {Fuller}(1984)}]{chow1984response}%
  \BibitemOpen
  \bibfield  {author} {\bibinfo {author} {\bibfnamefont {A.~W.}\ \bibnamefont {Chow}}\ and\ \bibinfo {author} {\bibfnamefont {G.~G.}\ \bibnamefont {Fuller}},\ }\bibfield  {title} {\enquote {\bibinfo {title} {Response of moderately concentrated xanthan gum solutions to time-dependent flows using two-color flow birefringence},}\ }\href@noop {} {\bibfield  {journal} {\bibinfo  {journal} {Journal of Rheology}\ }\textbf {\bibinfo {volume} {28}},\ \bibinfo {pages} {23--43} (\bibinfo {year} {1984})}\BibitemShut {NoStop}%
\bibitem [{\citenamefont {Meyer}\ \emph {et~al.}(1993)\citenamefont {Meyer}, \citenamefont {Fuller}, \citenamefont {Clark},\ and\ \citenamefont {Kulicke}}]{meyer1993investigation}%
  \BibitemOpen
  \bibfield  {author} {\bibinfo {author} {\bibfnamefont {E.~L.}\ \bibnamefont {Meyer}}, \bibinfo {author} {\bibfnamefont {G.~G.}\ \bibnamefont {Fuller}}, \bibinfo {author} {\bibfnamefont {R.~C.}\ \bibnamefont {Clark}}, \ and\ \bibinfo {author} {\bibfnamefont {W.}~\bibnamefont {Kulicke}},\ }\bibfield  {title} {\enquote {\bibinfo {title} {Investigation of xanthan gum solution behavior under shear flow using rheooptical techniques},}\ }\href@noop {} {\bibfield  {journal} {\bibinfo  {journal} {Macromolecules}\ }\textbf {\bibinfo {volume} {26}},\ \bibinfo {pages} {504--511} (\bibinfo {year} {1993})}\BibitemShut {NoStop}%
\bibitem [{\citenamefont {Humbert}\ and\ \citenamefont {Decruppe}(1998{\natexlab{a}})}]{humbert1998flow}%
  \BibitemOpen
  \bibfield  {author} {\bibinfo {author} {\bibfnamefont {C.}~\bibnamefont {Humbert}}\ and\ \bibinfo {author} {\bibfnamefont {J.}~\bibnamefont {Decruppe}},\ }\bibfield  {title} {\enquote {\bibinfo {title} {Flow birefringence and stress optical law of viscoelastic solutions of cationic surfactants and sodium salicylate},}\ }\href@noop {} {\bibfield  {journal} {\bibinfo  {journal} {The European Physical Journal B-Condensed Matter and Complex Systems}\ }\textbf {\bibinfo {volume} {6}},\ \bibinfo {pages} {511--518} (\bibinfo {year} {1998}{\natexlab{a}})}\BibitemShut {NoStop}%
\bibitem [{\citenamefont {Sridhar}, \citenamefont {Nguyen},\ and\ \citenamefont {Fuller}(2000)}]{sridhar2000birefringence}%
  \BibitemOpen
  \bibfield  {author} {\bibinfo {author} {\bibfnamefont {T.}~\bibnamefont {Sridhar}}, \bibinfo {author} {\bibfnamefont {D.~A.}\ \bibnamefont {Nguyen}}, \ and\ \bibinfo {author} {\bibfnamefont {G.}~\bibnamefont {Fuller}},\ }\bibfield  {title} {\enquote {\bibinfo {title} {Birefringence and stress growth in uniaxial extension of polymer solutions},}\ }\href@noop {} {\bibfield  {journal} {\bibinfo  {journal} {Journal of Non-Newtonian Fluid Mechanics}\ }\textbf {\bibinfo {volume} {90}},\ \bibinfo {pages} {299--315} (\bibinfo {year} {2000})}\BibitemShut {NoStop}%
\bibitem [{\citenamefont {Haward}, \citenamefont {Sharma},\ and\ \citenamefont {Odell}(2011)}]{haward2011extensional}%
  \BibitemOpen
  \bibfield  {author} {\bibinfo {author} {\bibfnamefont {S.~J.}\ \bibnamefont {Haward}}, \bibinfo {author} {\bibfnamefont {V.}~\bibnamefont {Sharma}}, \ and\ \bibinfo {author} {\bibfnamefont {J.~A.}\ \bibnamefont {Odell}},\ }\bibfield  {title} {\enquote {\bibinfo {title} {Extensional opto-rheometry with biofluids and ultra-dilute polymer solutions},}\ }\href@noop {} {\bibfield  {journal} {\bibinfo  {journal} {Soft Matter}\ }\textbf {\bibinfo {volume} {7}},\ \bibinfo {pages} {9908--9921} (\bibinfo {year} {2011})}\BibitemShut {NoStop}%
\bibitem [{\citenamefont {Haward}\ \emph {et~al.}(2012)\citenamefont {Haward}, \citenamefont {Ober}, \citenamefont {Oliveira}, \citenamefont {Alves},\ and\ \citenamefont {McKinley}}]{haward2012extensional}%
  \BibitemOpen
  \bibfield  {author} {\bibinfo {author} {\bibfnamefont {S.~J.}\ \bibnamefont {Haward}}, \bibinfo {author} {\bibfnamefont {T.~J.}\ \bibnamefont {Ober}}, \bibinfo {author} {\bibfnamefont {M.~S.}\ \bibnamefont {Oliveira}}, \bibinfo {author} {\bibfnamefont {M.~A.}\ \bibnamefont {Alves}}, \ and\ \bibinfo {author} {\bibfnamefont {G.~H.}\ \bibnamefont {McKinley}},\ }\bibfield  {title} {\enquote {\bibinfo {title} {Extensional rheology and elastic instabilities of a wormlike micellar solution in a microfluidic cross-slot device},}\ }\href@noop {} {\bibfield  {journal} {\bibinfo  {journal} {Soft Matter}\ }\textbf {\bibinfo {volume} {8}},\ \bibinfo {pages} {536--555} (\bibinfo {year} {2012})}\BibitemShut {NoStop}%
\bibitem [{\citenamefont {Pathak}\ and\ \citenamefont {Hudson}(2006)}]{pathak2006rheo}%
  \BibitemOpen
  \bibfield  {author} {\bibinfo {author} {\bibfnamefont {J.~A.}\ \bibnamefont {Pathak}}\ and\ \bibinfo {author} {\bibfnamefont {S.~D.}\ \bibnamefont {Hudson}},\ }\bibfield  {title} {\enquote {\bibinfo {title} {Rheo-optics of equilibrium polymer solutions: Wormlike micelles in elongational flow in a microfluidic cross-slot},}\ }\href@noop {} {\bibfield  {journal} {\bibinfo  {journal} {Macromolecules}\ }\textbf {\bibinfo {volume} {39}},\ \bibinfo {pages} {8782--8792} (\bibinfo {year} {2006})}\BibitemShut {NoStop}%
\bibitem [{\citenamefont {Humbert}\ and\ \citenamefont {Decruppe}(1998{\natexlab{b}})}]{humbert1998stress}%
  \BibitemOpen
  \bibfield  {author} {\bibinfo {author} {\bibfnamefont {C.}~\bibnamefont {Humbert}}\ and\ \bibinfo {author} {\bibfnamefont {J.}~\bibnamefont {Decruppe}},\ }\bibfield  {title} {\enquote {\bibinfo {title} {Stress optical coefficient of viscoelastic solutions of cetyltrimethylammonium bromide and potassium bromide},}\ }\href@noop {} {\bibfield  {journal} {\bibinfo  {journal} {Colloid and Polymer Science}\ }\textbf {\bibinfo {volume} {276}},\ \bibinfo {pages} {160--168} (\bibinfo {year} {1998}{\natexlab{b}})}\BibitemShut {NoStop}%
\bibitem [{\citenamefont {Decruppe}\ and\ \citenamefont {Ponton}(2003)}]{decruppe2003flow}%
  \BibitemOpen
  \bibfield  {author} {\bibinfo {author} {\bibfnamefont {J.}~\bibnamefont {Decruppe}}\ and\ \bibinfo {author} {\bibfnamefont {A.}~\bibnamefont {Ponton}},\ }\bibfield  {title} {\enquote {\bibinfo {title} {Flow birefringence, stress optical rule and rheology of four micellar solutions with the same low shear viscosity},}\ }\href@noop {} {\bibfield  {journal} {\bibinfo  {journal} {The European Physical Journal E}\ }\textbf {\bibinfo {volume} {10}},\ \bibinfo {pages} {201--207} (\bibinfo {year} {2003})}\BibitemShut {NoStop}%
\bibitem [{\citenamefont {Ge}, \citenamefont {Shi},\ and\ \citenamefont {Zakin}(2012)}]{ge2012rheo}%
  \BibitemOpen
  \bibfield  {author} {\bibinfo {author} {\bibfnamefont {W.}~\bibnamefont {Ge}}, \bibinfo {author} {\bibfnamefont {H.}~\bibnamefont {Shi}}, \ and\ \bibinfo {author} {\bibfnamefont {J.~L.}\ \bibnamefont {Zakin}},\ }\bibfield  {title} {\enquote {\bibinfo {title} {Rheo-optics of cationic surfactant micellar solutions with mixed aromatic counterions},}\ }\href@noop {} {\bibfield  {journal} {\bibinfo  {journal} {Rheologica Acta}\ }\textbf {\bibinfo {volume} {51}},\ \bibinfo {pages} {249--258} (\bibinfo {year} {2012})}\BibitemShut {NoStop}%
\bibitem [{\citenamefont {Bass}(2010)}]{bass2010handbook}%
  \BibitemOpen
  \bibfield  {author} {\bibinfo {author} {\bibfnamefont {M.}~\bibnamefont {Bass}},\ }\href@noop {} {\emph {\bibinfo {title} {Handbook of Optics: Volume I. Geometrical and Physical Optics, Polarized Light, Components and Instruments}}}\ (\bibinfo  {publisher} {McGraw-Hill},\ \bibinfo {year} {2010})\BibitemShut {NoStop}%
\bibitem [{\citenamefont {Rastogi}(2015)}]{rastogi2015digital}%
  \BibitemOpen
  \bibfield  {author} {\bibinfo {author} {\bibfnamefont {P.}~\bibnamefont {Rastogi}},\ }\href@noop {} {\emph {\bibinfo {title} {Digital Optical Measurement Techniques and Applications}}}\ (\bibinfo  {publisher} {Artech House},\ \bibinfo {year} {2015})\BibitemShut {NoStop}%
\bibitem [{\citenamefont {Ito}, \citenamefont {Yoshitake},\ and\ \citenamefont {Takahashi}(2015)}]{ito2015temporal}%
  \BibitemOpen
  \bibfield  {author} {\bibinfo {author} {\bibfnamefont {M.}~\bibnamefont {Ito}}, \bibinfo {author} {\bibfnamefont {Y.}~\bibnamefont {Yoshitake}}, \ and\ \bibinfo {author} {\bibfnamefont {T.}~\bibnamefont {Takahashi}},\ }\bibfield  {title} {\enquote {\bibinfo {title} {Temporal shear oscillation in steady shear flow of ctab/nasal wormlike micellar solutions},}\ }\href@noop {} {\bibfield  {journal} {\bibinfo  {journal} {Journal of the Society of Rheology, Japan (in Japanese)}\ }\textbf {\bibinfo {volume} {43}},\ \bibinfo {pages} {39--45} (\bibinfo {year} {2015})}\BibitemShut {NoStop}%
\bibitem [{\citenamefont {Ito}, \citenamefont {Yoshitake},\ and\ \citenamefont {Takahashi}(2016)}]{ito2016shear}%
  \BibitemOpen
  \bibfield  {author} {\bibinfo {author} {\bibfnamefont {M.}~\bibnamefont {Ito}}, \bibinfo {author} {\bibfnamefont {Y.}~\bibnamefont {Yoshitake}}, \ and\ \bibinfo {author} {\bibfnamefont {T.}~\bibnamefont {Takahashi}},\ }\bibfield  {title} {\enquote {\bibinfo {title} {Shear-induced structure change in shear-banding of a wormlike micellar solution in concentric cylinder flow},}\ }\href@noop {} {\bibfield  {journal} {\bibinfo  {journal} {Journal of Rheology}\ }\textbf {\bibinfo {volume} {60}},\ \bibinfo {pages} {1019--1029} (\bibinfo {year} {2016})}\BibitemShut {NoStop}%
\bibitem [{\citenamefont {Worby}\ \emph {et~al.}(2024)\citenamefont {Worby}, \citenamefont {Nakamine}, \citenamefont {Yokoyama}, \citenamefont {Muto},\ and\ \citenamefont {Tagawa}}]{worby2024examination}%
  \BibitemOpen
  \bibfield  {author} {\bibinfo {author} {\bibfnamefont {W.~K.~A.}\ \bibnamefont {Worby}}, \bibinfo {author} {\bibfnamefont {K.}~\bibnamefont {Nakamine}}, \bibinfo {author} {\bibfnamefont {Y.}~\bibnamefont {Yokoyama}}, \bibinfo {author} {\bibfnamefont {M.}~\bibnamefont {Muto}}, \ and\ \bibinfo {author} {\bibfnamefont {Y.}~\bibnamefont {Tagawa}},\ }\bibfield  {title} {\enquote {\bibinfo {title} {Examination of flow birefringence induced by the shear components along the optical axis using a parallel-plate-type rheometer},}\ }\href@noop {} {\bibfield  {journal} {\bibinfo  {journal} {Scientific Reports}\ }\textbf {\bibinfo {volume} {14}},\ \bibinfo {pages} {21931} (\bibinfo {year} {2024})}\BibitemShut {NoStop}%
\bibitem [{\citenamefont {Yamamoto}\ \emph {et~al.}(2024)\citenamefont {Yamamoto}, \citenamefont {Yamagata}, \citenamefont {Sato}, \citenamefont {Nakamura}, \citenamefont {Sato}, \citenamefont {Naito}, \citenamefont {Chung},\ and\ \citenamefont {Katashima}}]{yamamoto2024elucidating}%
  \BibitemOpen
  \bibfield  {author} {\bibinfo {author} {\bibfnamefont {Y.}~\bibnamefont {Yamamoto}}, \bibinfo {author} {\bibfnamefont {Y.}~\bibnamefont {Yamagata}}, \bibinfo {author} {\bibfnamefont {T.}~\bibnamefont {Sato}}, \bibinfo {author} {\bibfnamefont {K.}~\bibnamefont {Nakamura}}, \bibinfo {author} {\bibfnamefont {R.}~\bibnamefont {Sato}}, \bibinfo {author} {\bibfnamefont {M.}~\bibnamefont {Naito}}, \bibinfo {author} {\bibfnamefont {U.-i.}\ \bibnamefont {Chung}}, \ and\ \bibinfo {author} {\bibfnamefont {T.}~\bibnamefont {Katashima}},\ }\bibfield  {title} {\enquote {\bibinfo {title} {Elucidating nonlinear stress relaxation in transient networks through two-dimensional rheo-optics},}\ }\href@noop {} {\bibfield  {journal} {\bibinfo  {journal} {ACS Macro Letters}\ }\textbf {\bibinfo {volume} {13}},\ \bibinfo {pages} {1171--1178} (\bibinfo {year} {2024})}\BibitemShut {NoStop}%
\bibitem [{\citenamefont {Sato}\ \emph {et~al.}(2024)\citenamefont {Sato}, \citenamefont {Yamagata}, \citenamefont {Sato}, \citenamefont {Onuma}, \citenamefont {Miyamoto},\ and\ \citenamefont {Takahashi}}]{sato2024two}%
  \BibitemOpen
  \bibfield  {author} {\bibinfo {author} {\bibfnamefont {T.}~\bibnamefont {Sato}}, \bibinfo {author} {\bibfnamefont {Y.}~\bibnamefont {Yamagata}}, \bibinfo {author} {\bibfnamefont {Y.}~\bibnamefont {Sato}}, \bibinfo {author} {\bibfnamefont {T.}~\bibnamefont {Onuma}}, \bibinfo {author} {\bibfnamefont {K.}~\bibnamefont {Miyamoto}}, \ and\ \bibinfo {author} {\bibfnamefont {T.}~\bibnamefont {Takahashi}},\ }\bibfield  {title} {\enquote {\bibinfo {title} {Two-dimensional rheo-optical measurement system to study dynamics and structure of complex fluids},}\ }\href@noop {} {\bibfield  {journal} {\bibinfo  {journal} {Applied Rheology}\ }\textbf {\bibinfo {volume} {34}},\ \bibinfo {pages} {20240006} (\bibinfo {year} {2024})}\BibitemShut {NoStop}%
\bibitem [{\citenamefont {McKelvey}(1962)}]{mckelvey1962polymer}%
  \BibitemOpen
  \bibfield  {author} {\bibinfo {author} {\bibfnamefont {J.~M.}\ \bibnamefont {McKelvey}},\ }\href@noop {} {\emph {\bibinfo {title} {Polymer Processing}}}\ (\bibinfo  {publisher} {John Wiley \& Sons},\ \bibinfo {year} {1962})\BibitemShut {NoStop}%
\bibitem [{\citenamefont {Angstadt}\ and\ \citenamefont {Coulter}(2006)}]{angstadt2006investigation}%
  \BibitemOpen
  \bibfield  {author} {\bibinfo {author} {\bibfnamefont {D.~C.}\ \bibnamefont {Angstadt}}\ and\ \bibinfo {author} {\bibfnamefont {J.~P.}\ \bibnamefont {Coulter}},\ }\bibfield  {title} {\enquote {\bibinfo {title} {Investigation of melt manipulation phenomena during injection molding via in situ birefringence observation},}\ }\href@noop {} {\bibfield  {journal} {\bibinfo  {journal} {Polymer Engineering \& Science}\ }\textbf {\bibinfo {volume} {46}},\ \bibinfo {pages} {1691--1697} (\bibinfo {year} {2006})}\BibitemShut {NoStop}%
\bibitem [{\citenamefont {Kim}\ \emph {et~al.}(2005)\citenamefont {Kim}, \citenamefont {Isayev}, \citenamefont {Kwon},\ and\ \citenamefont {Van~Sweden}}]{kim2005modeling}%
  \BibitemOpen
  \bibfield  {author} {\bibinfo {author} {\bibfnamefont {K.~H.}\ \bibnamefont {Kim}}, \bibinfo {author} {\bibfnamefont {A.}~\bibnamefont {Isayev}}, \bibinfo {author} {\bibfnamefont {K.}~\bibnamefont {Kwon}}, \ and\ \bibinfo {author} {\bibfnamefont {C.}~\bibnamefont {Van~Sweden}},\ }\bibfield  {title} {\enquote {\bibinfo {title} {Modeling and experimental study of birefringence in injection molding of semicrystalline polymers},}\ }\href@noop {} {\bibfield  {journal} {\bibinfo  {journal} {Polymer}\ }\textbf {\bibinfo {volume} {46}},\ \bibinfo {pages} {4183--4203} (\bibinfo {year} {2005})}\BibitemShut {NoStop}%
\bibitem [{\citenamefont {Min}\ and\ \citenamefont {Yoon}(2011)}]{min2011experimental}%
  \BibitemOpen
  \bibfield  {author} {\bibinfo {author} {\bibfnamefont {I.}~\bibnamefont {Min}}\ and\ \bibinfo {author} {\bibfnamefont {K.}~\bibnamefont {Yoon}},\ }\bibfield  {title} {\enquote {\bibinfo {title} {An experimental study on the effects of injection-molding types for the birefringence distribution in polycarbonate discs},}\ }\href@noop {} {\bibfield  {journal} {\bibinfo  {journal} {Korea-Australia Rheology Journal}\ }\textbf {\bibinfo {volume} {23}},\ \bibinfo {pages} {155--162} (\bibinfo {year} {2011})}\BibitemShut {NoStop}%
\bibitem [{\citenamefont {Lin}\ and\ \citenamefont {Chen}(2019)}]{lin2019grey}%
  \BibitemOpen
  \bibfield  {author} {\bibinfo {author} {\bibfnamefont {C.-M.}\ \bibnamefont {Lin}}\ and\ \bibinfo {author} {\bibfnamefont {Y.-W.}\ \bibnamefont {Chen}},\ }\bibfield  {title} {\enquote {\bibinfo {title} {Grey optimization of injection molding processing of plastic optical lens based on joint consideration of aberration and birefringence effects},}\ }\href@noop {} {\bibfield  {journal} {\bibinfo  {journal} {Microsystem Technologies}\ }\textbf {\bibinfo {volume} {25}},\ \bibinfo {pages} {621--631} (\bibinfo {year} {2019})}\BibitemShut {NoStop}%
\bibitem [{\citenamefont {Noto}\ \emph {et~al.}(2020)\citenamefont {Noto}, \citenamefont {Tasaka}, \citenamefont {Hitomi},\ and\ \citenamefont {Murai}}]{noto2020applicability}%
  \BibitemOpen
  \bibfield  {author} {\bibinfo {author} {\bibfnamefont {D.}~\bibnamefont {Noto}}, \bibinfo {author} {\bibfnamefont {Y.}~\bibnamefont {Tasaka}}, \bibinfo {author} {\bibfnamefont {J.}~\bibnamefont {Hitomi}}, \ and\ \bibinfo {author} {\bibfnamefont {Y.}~\bibnamefont {Murai}},\ }\bibfield  {title} {\enquote {\bibinfo {title} {Applicability evaluation of the stress-optic law in newtonian fluids toward stress field measurements},}\ }\href@noop {} {\bibfield  {journal} {\bibinfo  {journal} {Physical Review Research}\ }\textbf {\bibinfo {volume} {2}},\ \bibinfo {pages} {043111} (\bibinfo {year} {2020})}\BibitemShut {NoStop}%
\bibitem [{\citenamefont {Rastogi}(2003)}]{rastogi2003photomechanics}%
  \BibitemOpen
  \bibfield  {author} {\bibinfo {author} {\bibfnamefont {P.~K.}\ \bibnamefont {Rastogi}},\ }\href@noop {} {\emph {\bibinfo {title} {Photomechanics}}},\ Vol.~\bibinfo {volume} {77}\ (\bibinfo  {publisher} {Springer Science \& Business Media},\ \bibinfo {year} {2003})\BibitemShut {NoStop}%
\bibitem [{\citenamefont {Rastogi}\ and\ \citenamefont {Hack}(2013)}]{rastogi2013optical}%
  \BibitemOpen
  \bibfield  {author} {\bibinfo {author} {\bibfnamefont {P.~K.}\ \bibnamefont {Rastogi}}\ and\ \bibinfo {author} {\bibfnamefont {E.}~\bibnamefont {Hack}},\ }\href@noop {} {\emph {\bibinfo {title} {Optical Methods for Solid Mechanics: A Full-Field Approach}}}\ (\bibinfo  {publisher} {John Wiley \& Sons},\ \bibinfo {year} {2013})\BibitemShut {NoStop}%
\bibitem [{\citenamefont {Ebisawa}, \citenamefont {Otani},\ and\ \citenamefont {Umeda}(2007)}]{ebisawa2007mechanical}%
  \BibitemOpen
  \bibfield  {author} {\bibinfo {author} {\bibfnamefont {M.}~\bibnamefont {Ebisawa}}, \bibinfo {author} {\bibfnamefont {Y.}~\bibnamefont {Otani}}, \ and\ \bibinfo {author} {\bibfnamefont {N.}~\bibnamefont {Umeda}},\ }\bibfield  {title} {\enquote {\bibinfo {title} {Mechanical characterization measurement of polymer material under stress by birefringence microscope},}\ }\href@noop {} {\bibfield  {journal} {\bibinfo  {journal} {Optical Review}\ }\textbf {\bibinfo {volume} {14}},\ \bibinfo {pages} {310--313} (\bibinfo {year} {2007})}\BibitemShut {NoStop}%
\bibitem [{\citenamefont {Sun}\ \emph {et~al.}(2023)\citenamefont {Sun}, \citenamefont {Wang}, \citenamefont {Li}, \citenamefont {Huo}, \citenamefont {Wang}, \citenamefont {Li},\ and\ \citenamefont {Wang}}]{sun2023measurement}%
  \BibitemOpen
  \bibfield  {author} {\bibinfo {author} {\bibfnamefont {X.}~\bibnamefont {Sun}}, \bibinfo {author} {\bibfnamefont {S.}~\bibnamefont {Wang}}, \bibinfo {author} {\bibfnamefont {L.}~\bibnamefont {Li}}, \bibinfo {author} {\bibfnamefont {Z.}~\bibnamefont {Huo}}, \bibinfo {author} {\bibfnamefont {L.}~\bibnamefont {Wang}}, \bibinfo {author} {\bibfnamefont {C.}~\bibnamefont {Li}}, \ and\ \bibinfo {author} {\bibfnamefont {Z.}~\bibnamefont {Wang}},\ }\bibfield  {title} {\enquote {\bibinfo {title} {Measurement of stress-optic coefficients for metals in the visible to near-infrared spectrum with spectroscopic ellipsometry},}\ }\href@noop {} {\bibfield  {journal} {\bibinfo  {journal} {Optics and Lasers in Engineering}\ }\textbf {\bibinfo {volume} {161}},\ \bibinfo {pages} {107362} (\bibinfo {year} {2023})}\BibitemShut {NoStop}%
\bibitem [{\citenamefont {Talbott}\ and\ \citenamefont {Goddard}(1979)}]{talbott1979streaming}%
  \BibitemOpen
  \bibfield  {author} {\bibinfo {author} {\bibfnamefont {W.~H.}\ \bibnamefont {Talbott}}\ and\ \bibinfo {author} {\bibfnamefont {J.~D.}\ \bibnamefont {Goddard}},\ }\bibfield  {title} {\enquote {\bibinfo {title} {Streaming birefringence in extensional flow of polymer solutions},}\ }\href@noop {} {\bibfield  {journal} {\bibinfo  {journal} {Rheologica Acta}\ }\textbf {\bibinfo {volume} {18}},\ \bibinfo {pages} {505--517} (\bibinfo {year} {1979})}\BibitemShut {NoStop}%
\bibitem [{\citenamefont {Doyle}\ \emph {et~al.}(1998)\citenamefont {Doyle}, \citenamefont {Shaqfeh}, \citenamefont {McKinley},\ and\ \citenamefont {Spiegelberg}}]{doyle1998relaxation}%
  \BibitemOpen
  \bibfield  {author} {\bibinfo {author} {\bibfnamefont {P.~S.}\ \bibnamefont {Doyle}}, \bibinfo {author} {\bibfnamefont {E.~S.}\ \bibnamefont {Shaqfeh}}, \bibinfo {author} {\bibfnamefont {G.~H.}\ \bibnamefont {McKinley}}, \ and\ \bibinfo {author} {\bibfnamefont {S.~H.}\ \bibnamefont {Spiegelberg}},\ }\bibfield  {title} {\enquote {\bibinfo {title} {Relaxation of dilute polymer solutions following extensional flow},}\ }\href@noop {} {\bibfield  {journal} {\bibinfo  {journal} {Journal of Non-Newtonian Fluid Mechanics}\ }\textbf {\bibinfo {volume} {76}},\ \bibinfo {pages} {79--110} (\bibinfo {year} {1998})}\BibitemShut {NoStop}%
\bibitem [{\citenamefont {Janeschitz-Kriegl}(2012)}]{janeschitz2012polymer}%
  \BibitemOpen
  \bibfield  {author} {\bibinfo {author} {\bibfnamefont {H.}~\bibnamefont {Janeschitz-Kriegl}},\ }\href@noop {} {\emph {\bibinfo {title} {Polymer Melt Rheology and Flow Birefringence}}},\ Vol.~\bibinfo {volume} {6}\ (\bibinfo  {publisher} {Springer Science \& Business Media},\ \bibinfo {year} {2012})\BibitemShut {NoStop}%
\bibitem [{\citenamefont {Zhao}, \citenamefont {Shen},\ and\ \citenamefont {Haward}(2016)}]{zhao2016flow}%
  \BibitemOpen
  \bibfield  {author} {\bibinfo {author} {\bibfnamefont {Y.}~\bibnamefont {Zhao}}, \bibinfo {author} {\bibfnamefont {A.~Q.}\ \bibnamefont {Shen}}, \ and\ \bibinfo {author} {\bibfnamefont {S.~J.}\ \bibnamefont {Haward}},\ }\bibfield  {title} {\enquote {\bibinfo {title} {Flow of wormlike micellar solutions around confined microfluidic cylinders},}\ }\href@noop {} {\bibfield  {journal} {\bibinfo  {journal} {Soft Matter}\ }\textbf {\bibinfo {volume} {12}},\ \bibinfo {pages} {8666--8681} (\bibinfo {year} {2016})}\BibitemShut {NoStop}%
\bibitem [{\citenamefont {Iwata}\ \emph {et~al.}(2019)\citenamefont {Iwata}, \citenamefont {Takahashi}, \citenamefont {Onuma}, \citenamefont {Nagumo},\ and\ \citenamefont {Mori}}]{iwata2019local}%
  \BibitemOpen
  \bibfield  {author} {\bibinfo {author} {\bibfnamefont {S.}~\bibnamefont {Iwata}}, \bibinfo {author} {\bibfnamefont {T.}~\bibnamefont {Takahashi}}, \bibinfo {author} {\bibfnamefont {T.}~\bibnamefont {Onuma}}, \bibinfo {author} {\bibfnamefont {R.}~\bibnamefont {Nagumo}}, \ and\ \bibinfo {author} {\bibfnamefont {H.}~\bibnamefont {Mori}},\ }\bibfield  {title} {\enquote {\bibinfo {title} {Local flow around a tiny bubble under a pressure-oscillation field in a viscoelastic worm-like micellar solution},}\ }\href@noop {} {\bibfield  {journal} {\bibinfo  {journal} {Journal of Non-Newtonian Fluid Mechanics}\ }\textbf {\bibinfo {volume} {263}},\ \bibinfo {pages} {24--32} (\bibinfo {year} {2019})}\BibitemShut {NoStop}%
\bibitem [{\citenamefont {K{\'a}d{\'a}r}, \citenamefont {Spirk},\ and\ \citenamefont {Nypel{\"o}}(2021)}]{kadar2021cellulose}%
  \BibitemOpen
  \bibfield  {author} {\bibinfo {author} {\bibfnamefont {R.}~\bibnamefont {K{\'a}d{\'a}r}}, \bibinfo {author} {\bibfnamefont {S.}~\bibnamefont {Spirk}}, \ and\ \bibinfo {author} {\bibfnamefont {T.}~\bibnamefont {Nypel{\"o}}},\ }\bibfield  {title} {\enquote {\bibinfo {title} {Cellulose nanocrystal liquid crystal phases: Progress and challenges in characterization using rheology coupled to optics, scattering, and spectroscopy},}\ }\href@noop {} {\bibfield  {journal} {\bibinfo  {journal} {ACS Nano}\ }\textbf {\bibinfo {volume} {15}},\ \bibinfo {pages} {7931--7945} (\bibinfo {year} {2021})}\BibitemShut {NoStop}%
\bibitem [{\citenamefont {Lane}, \citenamefont {Rode},\ and\ \citenamefont {R{\"o}sgen}(2022)}]{lane2022birefringent}%
  \BibitemOpen
  \bibfield  {author} {\bibinfo {author} {\bibfnamefont {C.}~\bibnamefont {Lane}}, \bibinfo {author} {\bibfnamefont {D.}~\bibnamefont {Rode}}, \ and\ \bibinfo {author} {\bibfnamefont {T.}~\bibnamefont {R{\"o}sgen}},\ }\bibfield  {title} {\enquote {\bibinfo {title} {Birefringent properties of aqueous cellulose nanocrystal suspensions},}\ }\href@noop {} {\bibfield  {journal} {\bibinfo  {journal} {Cellulose}\ }\textbf {\bibinfo {volume} {29}},\ \bibinfo {pages} {6093--6107} (\bibinfo {year} {2022})}\BibitemShut {NoStop}%
\bibitem [{\citenamefont {Shikata}, \citenamefont {Dahman},\ and\ \citenamefont {Pearson}(1994)}]{shikata1994rheo}%
  \BibitemOpen
  \bibfield  {author} {\bibinfo {author} {\bibfnamefont {T.}~\bibnamefont {Shikata}}, \bibinfo {author} {\bibfnamefont {S.~J.}\ \bibnamefont {Dahman}}, \ and\ \bibinfo {author} {\bibfnamefont {D.~S.}\ \bibnamefont {Pearson}},\ }\bibfield  {title} {\enquote {\bibinfo {title} {Rheo-optical behavior of wormlike micelles},}\ }\href@noop {} {\bibfield  {journal} {\bibinfo  {journal} {Langmuir}\ }\textbf {\bibinfo {volume} {10}},\ \bibinfo {pages} {3470--3476} (\bibinfo {year} {1994})}\BibitemShut {NoStop}%
\bibitem [{\citenamefont {Wheeler}, \citenamefont {Izu},\ and\ \citenamefont {Fuller}(1996)}]{wheeler1996structure}%
  \BibitemOpen
  \bibfield  {author} {\bibinfo {author} {\bibfnamefont {E.~K.}\ \bibnamefont {Wheeler}}, \bibinfo {author} {\bibfnamefont {P.}~\bibnamefont {Izu}}, \ and\ \bibinfo {author} {\bibfnamefont {G.~G.}\ \bibnamefont {Fuller}},\ }\bibfield  {title} {\enquote {\bibinfo {title} {Structure and rheology of wormlike micelles},}\ }\href@noop {} {\bibfield  {journal} {\bibinfo  {journal} {Rheologica Acta}\ }\textbf {\bibinfo {volume} {35}},\ \bibinfo {pages} {139--149} (\bibinfo {year} {1996})}\BibitemShut {NoStop}%
\bibitem [{\citenamefont {Takahashi}, \citenamefont {Sugata},\ and\ \citenamefont {Shirakashi}(2002)}]{C_Takahashi}%
  \BibitemOpen
  \bibfield  {author} {\bibinfo {author} {\bibfnamefont {T.}~\bibnamefont {Takahashi}}, \bibinfo {author} {\bibfnamefont {H.}~\bibnamefont {Sugata}}, \ and\ \bibinfo {author} {\bibfnamefont {M.}~\bibnamefont {Shirakashi}},\ }\bibfield  {title} {\enquote {\bibinfo {title} {Rheo-optic behavior of wormlike miicelles under a shear-indeuced structure formational condition},}\ }\href@noop {} {\bibfield  {journal} {\bibinfo  {journal} {Journal of the Society of Rheology, Japan (in Japanese)}\ }\textbf {\bibinfo {volume} {30}},\ \bibinfo {pages} {109--113} (\bibinfo {year} {2002})}\BibitemShut {NoStop}%
\bibitem [{\citenamefont {Haward}\ \emph {et~al.}(2019)\citenamefont {Haward}, \citenamefont {Kitajima}, \citenamefont {Toda-Peters}, \citenamefont {Takahashi},\ and\ \citenamefont {Shen}}]{haward2019flow}%
  \BibitemOpen
  \bibfield  {author} {\bibinfo {author} {\bibfnamefont {S.~J.}\ \bibnamefont {Haward}}, \bibinfo {author} {\bibfnamefont {N.}~\bibnamefont {Kitajima}}, \bibinfo {author} {\bibfnamefont {K.}~\bibnamefont {Toda-Peters}}, \bibinfo {author} {\bibfnamefont {T.}~\bibnamefont {Takahashi}}, \ and\ \bibinfo {author} {\bibfnamefont {A.~Q.}\ \bibnamefont {Shen}},\ }\bibfield  {title} {\enquote {\bibinfo {title} {Flow of wormlike micellar solutions around microfluidic cylinders with high aspect ratio and low blockage ratio},}\ }\href@noop {} {\bibfield  {journal} {\bibinfo  {journal} {Soft Matter}\ }\textbf {\bibinfo {volume} {15}},\ \bibinfo {pages} {1927--1941} (\bibinfo {year} {2019})}\BibitemShut {NoStop}%
\bibitem [{\citenamefont {Cardiel}\ \emph {et~al.}(2016)\citenamefont {Cardiel}, \citenamefont {Takagi}, \citenamefont {Tsai},\ and\ \citenamefont {Shen}}]{cardiel2016formation}%
  \BibitemOpen
  \bibfield  {author} {\bibinfo {author} {\bibfnamefont {J.~J.}\ \bibnamefont {Cardiel}}, \bibinfo {author} {\bibfnamefont {D.}~\bibnamefont {Takagi}}, \bibinfo {author} {\bibfnamefont {H.-F.}\ \bibnamefont {Tsai}}, \ and\ \bibinfo {author} {\bibfnamefont {A.~Q.}\ \bibnamefont {Shen}},\ }\bibfield  {title} {\enquote {\bibinfo {title} {Formation and flow behavior of micellar membranes in a {T}-shaped microchannel},}\ }\href@noop {} {\bibfield  {journal} {\bibinfo  {journal} {Soft Matter}\ }\textbf {\bibinfo {volume} {12}},\ \bibinfo {pages} {8226--8234} (\bibinfo {year} {2016})}\BibitemShut {NoStop}%
\bibitem [{\citenamefont {Shikata}, \citenamefont {Hirata},\ and\ \citenamefont {Kotaka}(1987)}]{shikata1987micelle}%
  \BibitemOpen
  \bibfield  {author} {\bibinfo {author} {\bibfnamefont {T.}~\bibnamefont {Shikata}}, \bibinfo {author} {\bibfnamefont {H.}~\bibnamefont {Hirata}}, \ and\ \bibinfo {author} {\bibfnamefont {T.}~\bibnamefont {Kotaka}},\ }\bibfield  {title} {\enquote {\bibinfo {title} {Micelle formation of detergent molecules in aqueous media: viscoelastic properties of aqueous cetyltrimethylammonium bromide solutions},}\ }\href@noop {} {\bibfield  {journal} {\bibinfo  {journal} {Langmuir}\ }\textbf {\bibinfo {volume} {3}},\ \bibinfo {pages} {1081--1086} (\bibinfo {year} {1987})}\BibitemShut {NoStop}%
\bibitem [{\citenamefont {Shikata}, \citenamefont {Hirata},\ and\ \citenamefont {Kotaka}(1988)}]{shikata1988micelle}%
  \BibitemOpen
  \bibfield  {author} {\bibinfo {author} {\bibfnamefont {T.}~\bibnamefont {Shikata}}, \bibinfo {author} {\bibfnamefont {H.}~\bibnamefont {Hirata}}, \ and\ \bibinfo {author} {\bibfnamefont {T.}~\bibnamefont {Kotaka}},\ }\bibfield  {title} {\enquote {\bibinfo {title} {Micelle formation of detergent molecules in aqueous media. 2. {R}ole of free salicylate ions on viscoelastic properties of aqueous cetyltrimethylammonium bromide-sodium salicylate solutions},}\ }\href@noop {} {\bibfield  {journal} {\bibinfo  {journal} {Langmuir}\ }\textbf {\bibinfo {volume} {4}},\ \bibinfo {pages} {354--359} (\bibinfo {year} {1988})}\BibitemShut {NoStop}%
\bibitem [{\citenamefont {Shikata}, \citenamefont {Hirata},\ and\ \citenamefont {Kotaka}(1989)}]{shikata1989micelle}%
  \BibitemOpen
  \bibfield  {author} {\bibinfo {author} {\bibfnamefont {T.}~\bibnamefont {Shikata}}, \bibinfo {author} {\bibfnamefont {H.}~\bibnamefont {Hirata}}, \ and\ \bibinfo {author} {\bibfnamefont {T.}~\bibnamefont {Kotaka}},\ }\bibfield  {title} {\enquote {\bibinfo {title} {Micelle formation of detergent molecules in aqueous media. 3. viscoelastic properties of aqueous cetyltrimethylammonium bromide-salicylic acid solutions},}\ }\href@noop {} {\bibfield  {journal} {\bibinfo  {journal} {Langmuir}\ }\textbf {\bibinfo {volume} {5}},\ \bibinfo {pages} {398--405} (\bibinfo {year} {1989})}\BibitemShut {NoStop}%
\bibitem [{\citenamefont {Shikata}\ \emph {et~al.}(1988)\citenamefont {Shikata}, \citenamefont {Hirata}, \citenamefont {Takatori},\ and\ \citenamefont {Osaki}}]{shikata1988nonlinear}%
  \BibitemOpen
  \bibfield  {author} {\bibinfo {author} {\bibfnamefont {T.}~\bibnamefont {Shikata}}, \bibinfo {author} {\bibfnamefont {H.}~\bibnamefont {Hirata}}, \bibinfo {author} {\bibfnamefont {E.}~\bibnamefont {Takatori}}, \ and\ \bibinfo {author} {\bibfnamefont {K.}~\bibnamefont {Osaki}},\ }\bibfield  {title} {\enquote {\bibinfo {title} {Nonlinear viscoelastic behavior of aqueous detergent solutions},}\ }\href@noop {} {\bibfield  {journal} {\bibinfo  {journal} {Journal of Non-Newtonian Fluid Mechanics}\ }\textbf {\bibinfo {volume} {28}},\ \bibinfo {pages} {171--182} (\bibinfo {year} {1988})}\BibitemShut {NoStop}%
\bibitem [{\citenamefont {Lodge}(1955)}]{lodge1955variation}%
  \BibitemOpen
  \bibfield  {author} {\bibinfo {author} {\bibfnamefont {A.}~\bibnamefont {Lodge}},\ }\bibfield  {title} {\enquote {\bibinfo {title} {Variation of flow birefringence with stress},}\ }\href@noop {} {\bibfield  {journal} {\bibinfo  {journal} {Nature}\ }\textbf {\bibinfo {volume} {176}},\ \bibinfo {pages} {838--839} (\bibinfo {year} {1955})}\BibitemShut {NoStop}%
\bibitem [{\citenamefont {Rothstein}\ and\ \citenamefont {McKinley}(2002{\natexlab{a}})}]{rothstein2002comparison}%
  \BibitemOpen
  \bibfield  {author} {\bibinfo {author} {\bibfnamefont {J.~P.}\ \bibnamefont {Rothstein}}\ and\ \bibinfo {author} {\bibfnamefont {G.~H.}\ \bibnamefont {McKinley}},\ }\bibfield  {title} {\enquote {\bibinfo {title} {A comparison of the stress and birefringence growth of dilute, semi-dilute and concentrated polymer solutions in uniaxial extensional flows},}\ }\href@noop {} {\bibfield  {journal} {\bibinfo  {journal} {Journal of Non-Newtonian Fluid Mechanics}\ }\textbf {\bibinfo {volume} {108}},\ \bibinfo {pages} {275--290} (\bibinfo {year} {2002}{\natexlab{a}})}\BibitemShut {NoStop}%
\bibitem [{\citenamefont {Rothstein}\ and\ \citenamefont {McKinley}(2002{\natexlab{b}})}]{rothstein2002inhomogeneous}%
  \BibitemOpen
  \bibfield  {author} {\bibinfo {author} {\bibfnamefont {J.~P.}\ \bibnamefont {Rothstein}}\ and\ \bibinfo {author} {\bibfnamefont {G.~H.}\ \bibnamefont {McKinley}},\ }\bibfield  {title} {\enquote {\bibinfo {title} {Inhomogeneous transient uniaxial extensional rheometry},}\ }\href@noop {} {\bibfield  {journal} {\bibinfo  {journal} {Journal of Rheology}\ }\textbf {\bibinfo {volume} {46}},\ \bibinfo {pages} {1419--1443} (\bibinfo {year} {2002}{\natexlab{b}})}\BibitemShut {NoStop}%
\bibitem [{\citenamefont {Rothstein}(2003)}]{rothstein2003transient}%
  \BibitemOpen
  \bibfield  {author} {\bibinfo {author} {\bibfnamefont {J.~P.}\ \bibnamefont {Rothstein}},\ }\bibfield  {title} {\enquote {\bibinfo {title} {Transient extensional rheology of wormlike micelle solutions},}\ }\href@noop {} {\bibfield  {journal} {\bibinfo  {journal} {Journal of Rheology}\ }\textbf {\bibinfo {volume} {47}},\ \bibinfo {pages} {1227--1247} (\bibinfo {year} {2003})}\BibitemShut {NoStop}%
\bibitem [{\citenamefont {Muto}\ \emph {et~al.}(2023)\citenamefont {Muto}, \citenamefont {Kikuchi}, \citenamefont {Yoshino}, \citenamefont {Muraoka}, \citenamefont {Iwata}, \citenamefont {Nakamura}, \citenamefont {Osuka},\ and\ \citenamefont {Tamano}}]{muto2023rheological}%
  \BibitemOpen
  \bibfield  {author} {\bibinfo {author} {\bibfnamefont {M.}~\bibnamefont {Muto}}, \bibinfo {author} {\bibfnamefont {K.}~\bibnamefont {Kikuchi}}, \bibinfo {author} {\bibfnamefont {T.}~\bibnamefont {Yoshino}}, \bibinfo {author} {\bibfnamefont {A.}~\bibnamefont {Muraoka}}, \bibinfo {author} {\bibfnamefont {S.}~\bibnamefont {Iwata}}, \bibinfo {author} {\bibfnamefont {M.}~\bibnamefont {Nakamura}}, \bibinfo {author} {\bibfnamefont {S.}~\bibnamefont {Osuka}}, \ and\ \bibinfo {author} {\bibfnamefont {S.}~\bibnamefont {Tamano}},\ }\bibfield  {title} {\enquote {\bibinfo {title} {Rheological characterization of human follicular fluid under shear and extensional stress conditions},}\ }\href@noop {} {\bibfield  {journal} {\bibinfo  {journal} {Frontiers in Physics}\ }\textbf {\bibinfo {volume} {11}},\ \bibinfo {pages} {1308322} (\bibinfo {year} {2023})}\BibitemShut {NoStop}%
\bibitem [{\citenamefont {Tamano}, \citenamefont {Ohashi},\ and\ \citenamefont {Morinishi}(2017)}]{tamano2017dynamics}%
  \BibitemOpen
  \bibfield  {author} {\bibinfo {author} {\bibfnamefont {S.}~\bibnamefont {Tamano}}, \bibinfo {author} {\bibfnamefont {Y.}~\bibnamefont {Ohashi}}, \ and\ \bibinfo {author} {\bibfnamefont {Y.}~\bibnamefont {Morinishi}},\ }\bibfield  {title} {\enquote {\bibinfo {title} {Dynamics of falling droplet and elongational properties of dilute nonionic surfactant solutions with drag-reducing ability},}\ }\href@noop {} {\bibfield  {journal} {\bibinfo  {journal} {Physics of Fluids}\ }\textbf {\bibinfo {volume} {29}} (\bibinfo {year} {2017})}\BibitemShut {NoStop}%
\bibitem [{\citenamefont {Amarouchene}\ \emph {et~al.}(2001)\citenamefont {Amarouchene}, \citenamefont {Bonn}, \citenamefont {Meunier},\ and\ \citenamefont {Kellay}}]{amarouchene2001inhibition}%
  \BibitemOpen
  \bibfield  {author} {\bibinfo {author} {\bibfnamefont {Y.}~\bibnamefont {Amarouchene}}, \bibinfo {author} {\bibfnamefont {D.}~\bibnamefont {Bonn}}, \bibinfo {author} {\bibfnamefont {J.}~\bibnamefont {Meunier}}, \ and\ \bibinfo {author} {\bibfnamefont {H.}~\bibnamefont {Kellay}},\ }\bibfield  {title} {\enquote {\bibinfo {title} {Inhibition of the finite-time singularity during droplet fission of a polymeric fluid},}\ }\href@noop {} {\bibfield  {journal} {\bibinfo  {journal} {Physical Review Letters}\ }\textbf {\bibinfo {volume} {86}},\ \bibinfo {pages} {3558} (\bibinfo {year} {2001})}\BibitemShut {NoStop}%
\bibitem [{\citenamefont {Onuma}\ and\ \citenamefont {Otani}(2014)}]{onuma2014development}%
  \BibitemOpen
  \bibfield  {author} {\bibinfo {author} {\bibfnamefont {T.}~\bibnamefont {Onuma}}\ and\ \bibinfo {author} {\bibfnamefont {Y.}~\bibnamefont {Otani}},\ }\bibfield  {title} {\enquote {\bibinfo {title} {A development of two-dimensional birefringence distribution measurement system with a sampling rate of 1.3 {MHz}},}\ }\href@noop {} {\bibfield  {journal} {\bibinfo  {journal} {Optics Communications}\ }\textbf {\bibinfo {volume} {315}},\ \bibinfo {pages} {69--73} (\bibinfo {year} {2014})}\BibitemShut {NoStop}%
\bibitem [{\citenamefont {Dinic}\ and\ \citenamefont {Sharma}(2020)}]{dinic2020flexibility}%
  \BibitemOpen
  \bibfield  {author} {\bibinfo {author} {\bibfnamefont {J.}~\bibnamefont {Dinic}}\ and\ \bibinfo {author} {\bibfnamefont {V.}~\bibnamefont {Sharma}},\ }\bibfield  {title} {\enquote {\bibinfo {title} {Flexibility, extensibility, and ratio of kuhn length to packing length govern the pinching dynamics, coil-stretch transition, and rheology of polymer solutions},}\ }\href@noop {} {\bibfield  {journal} {\bibinfo  {journal} {Macromolecules}\ }\textbf {\bibinfo {volume} {53}},\ \bibinfo {pages} {4821--4835} (\bibinfo {year} {2020})}\BibitemShut {NoStop}%
\bibitem [{\citenamefont {Dinic}, \citenamefont {Jimenez},\ and\ \citenamefont {Sharma}(2017)}]{dinic2017pinch}%
  \BibitemOpen
  \bibfield  {author} {\bibinfo {author} {\bibfnamefont {J.}~\bibnamefont {Dinic}}, \bibinfo {author} {\bibfnamefont {L.~N.}\ \bibnamefont {Jimenez}}, \ and\ \bibinfo {author} {\bibfnamefont {V.}~\bibnamefont {Sharma}},\ }\bibfield  {title} {\enquote {\bibinfo {title} {Pinch-off dynamics and dripping-onto-substrate ({DoS}) rheometry of complex fluids},}\ }\href@noop {} {\bibfield  {journal} {\bibinfo  {journal} {Lab on a Chip}\ }\textbf {\bibinfo {volume} {17}},\ \bibinfo {pages} {460--473} (\bibinfo {year} {2017})}\BibitemShut {NoStop}%
\bibitem [{\citenamefont {McKinley}\ and\ \citenamefont {Tripathi}(2000)}]{mckinley2000extract}%
  \BibitemOpen
  \bibfield  {author} {\bibinfo {author} {\bibfnamefont {G.~H.}\ \bibnamefont {McKinley}}\ and\ \bibinfo {author} {\bibfnamefont {A.}~\bibnamefont {Tripathi}},\ }\bibfield  {title} {\enquote {\bibinfo {title} {How to extract the {Newtonian} viscosity from capillary breakup measurements in a filament rheometer},}\ }\href@noop {} {\bibfield  {journal} {\bibinfo  {journal} {Journal of Rheology}\ }\textbf {\bibinfo {volume} {44}},\ \bibinfo {pages} {653--670} (\bibinfo {year} {2000})}\BibitemShut {NoStop}%
\bibitem [{\citenamefont {Clasen}\ \emph {et~al.}(2006)\citenamefont {Clasen}, \citenamefont {Plog}, \citenamefont {Kulicke}, \citenamefont {Owens}, \citenamefont {Macosko}, \citenamefont {Scriven}, \citenamefont {Verani},\ and\ \citenamefont {McKinley}}]{clasen2006dilute}%
  \BibitemOpen
  \bibfield  {author} {\bibinfo {author} {\bibfnamefont {C.}~\bibnamefont {Clasen}}, \bibinfo {author} {\bibfnamefont {J.~P.}\ \bibnamefont {Plog}}, \bibinfo {author} {\bibfnamefont {W.-M.}\ \bibnamefont {Kulicke}}, \bibinfo {author} {\bibfnamefont {M.}~\bibnamefont {Owens}}, \bibinfo {author} {\bibfnamefont {C.}~\bibnamefont {Macosko}}, \bibinfo {author} {\bibfnamefont {L.~E.}\ \bibnamefont {Scriven}}, \bibinfo {author} {\bibfnamefont {M.}~\bibnamefont {Verani}}, \ and\ \bibinfo {author} {\bibfnamefont {G.~H.}\ \bibnamefont {McKinley}},\ }\bibfield  {title} {\enquote {\bibinfo {title} {How dilute are dilute solutions in extensional flows?}}\ }\href@noop {} {\bibfield  {journal} {\bibinfo  {journal} {Journal of Rheology}\ }\textbf {\bibinfo {volume} {50}},\ \bibinfo {pages} {849--881} (\bibinfo {year} {2006})}\BibitemShut {NoStop}%
\bibitem [{\citenamefont {Dinic}\ and\ \citenamefont {Sharma}(2019)}]{dinic2019macromolecular}%
  \BibitemOpen
  \bibfield  {author} {\bibinfo {author} {\bibfnamefont {J.}~\bibnamefont {Dinic}}\ and\ \bibinfo {author} {\bibfnamefont {V.}~\bibnamefont {Sharma}},\ }\bibfield  {title} {\enquote {\bibinfo {title} {Macromolecular relaxation, strain, and extensibility determine elastocapillary thinning and extensional viscosity of polymer solutions},}\ }\href@noop {} {\bibfield  {journal} {\bibinfo  {journal} {Proceedings of the National Academy of Sciences}\ }\textbf {\bibinfo {volume} {116}},\ \bibinfo {pages} {8766--8774} (\bibinfo {year} {2019})}\BibitemShut {NoStop}%
\bibitem [{\citenamefont {Dinic}\ \emph {et~al.}(2015)\citenamefont {Dinic}, \citenamefont {Zhang}, \citenamefont {Jimenez},\ and\ \citenamefont {Sharma}}]{dinic2015extensional}%
  \BibitemOpen
  \bibfield  {author} {\bibinfo {author} {\bibfnamefont {J.}~\bibnamefont {Dinic}}, \bibinfo {author} {\bibfnamefont {Y.}~\bibnamefont {Zhang}}, \bibinfo {author} {\bibfnamefont {L.~N.}\ \bibnamefont {Jimenez}}, \ and\ \bibinfo {author} {\bibfnamefont {V.}~\bibnamefont {Sharma}},\ }\bibfield  {title} {\enquote {\bibinfo {title} {Extensional relaxation times of dilute, aqueous polymer solutions},}\ }\href@noop {} {\bibfield  {journal} {\bibinfo  {journal} {ACS Macro Letters}\ }\textbf {\bibinfo {volume} {4}},\ \bibinfo {pages} {804--808} (\bibinfo {year} {2015})}\BibitemShut {NoStop}%
\bibitem [{\citenamefont {Sur}\ and\ \citenamefont {Rothstein}(2018)}]{sur2018drop}%
  \BibitemOpen
  \bibfield  {author} {\bibinfo {author} {\bibfnamefont {S.}~\bibnamefont {Sur}}\ and\ \bibinfo {author} {\bibfnamefont {J.}~\bibnamefont {Rothstein}},\ }\bibfield  {title} {\enquote {\bibinfo {title} {Drop breakup dynamics of dilute polymer solutions: Effect of molecular weight, concentration, and viscosity},}\ }\href@noop {} {\bibfield  {journal} {\bibinfo  {journal} {Journal of Rheology}\ }\textbf {\bibinfo {volume} {62}},\ \bibinfo {pages} {1245--1259} (\bibinfo {year} {2018})}\BibitemShut {NoStop}%
\bibitem [{\citenamefont {McKinley}(2005)}]{mckinley2005visco}%
  \BibitemOpen
  \bibfield  {author} {\bibinfo {author} {\bibfnamefont {G.~H.}\ \bibnamefont {McKinley}},\ }\bibfield  {title} {\enquote {\bibinfo {title} {Visco-elasto-capillary thinning and break-up of complex fluids},}\ }\href@noop {} {\bibfield  {journal} {\bibinfo  {journal} {Annual Rheology Reviews}\ }\textbf {\bibinfo {volume} {1}} (\bibinfo {year} {2005})}\BibitemShut {NoStop}%
\bibitem [{\citenamefont {Omidvar}, \citenamefont {Wu},\ and\ \citenamefont {Mohammadigoushki}(2019)}]{omidvar2019detecting}%
  \BibitemOpen
  \bibfield  {author} {\bibinfo {author} {\bibfnamefont {R.}~\bibnamefont {Omidvar}}, \bibinfo {author} {\bibfnamefont {S.}~\bibnamefont {Wu}}, \ and\ \bibinfo {author} {\bibfnamefont {H.}~\bibnamefont {Mohammadigoushki}},\ }\bibfield  {title} {\enquote {\bibinfo {title} {Detecting wormlike micellar microstructure using extensional rheology},}\ }\href@noop {} {\bibfield  {journal} {\bibinfo  {journal} {Journal of Rheology}\ }\textbf {\bibinfo {volume} {63}},\ \bibinfo {pages} {33--44} (\bibinfo {year} {2019})}\BibitemShut {NoStop}%
\bibitem [{\citenamefont {Wagner}, \citenamefont {Bourouiba},\ and\ \citenamefont {McKinley}(2015)}]{wagner2015analytic}%
  \BibitemOpen
  \bibfield  {author} {\bibinfo {author} {\bibfnamefont {C.}~\bibnamefont {Wagner}}, \bibinfo {author} {\bibfnamefont {L.}~\bibnamefont {Bourouiba}}, \ and\ \bibinfo {author} {\bibfnamefont {G.~H.}\ \bibnamefont {McKinley}},\ }\bibfield  {title} {\enquote {\bibinfo {title} {An analytic solution for capillary thinning and breakup of {FENE-P} fluids},}\ }\href@noop {} {\bibfield  {journal} {\bibinfo  {journal} {Journal of Non-Newtonian Fluid Mechanics}\ }\textbf {\bibinfo {volume} {218}},\ \bibinfo {pages} {53--61} (\bibinfo {year} {2015})}\BibitemShut {NoStop}%
\bibitem [{\citenamefont {Prabhakar}, \citenamefont {Prakash},\ and\ \citenamefont {Sridhar}(2006)}]{prabhakar2006effect}%
  \BibitemOpen
  \bibfield  {author} {\bibinfo {author} {\bibfnamefont {R.}~\bibnamefont {Prabhakar}}, \bibinfo {author} {\bibfnamefont {J.~R.}\ \bibnamefont {Prakash}}, \ and\ \bibinfo {author} {\bibfnamefont {T.}~\bibnamefont {Sridhar}},\ }\bibfield  {title} {\enquote {\bibinfo {title} {Effect of configuration-dependent intramolecular hydrodynamic interaction on elastocapillary thinning and breakup of filaments of dilute polymer solutions},}\ }\href@noop {} {\bibfield  {journal} {\bibinfo  {journal} {Journal of Rheology}\ }\textbf {\bibinfo {volume} {50}},\ \bibinfo {pages} {925--947} (\bibinfo {year} {2006})}\BibitemShut {NoStop}%
\bibitem [{\citenamefont {Anna}\ and\ \citenamefont {McKinley}(2001)}]{Anna2001}%
  \BibitemOpen
  \bibfield  {author} {\bibinfo {author} {\bibfnamefont {S.~L.}\ \bibnamefont {Anna}}\ and\ \bibinfo {author} {\bibfnamefont {G.~H.}\ \bibnamefont {McKinley}},\ }\bibfield  {title} {\enquote {\bibinfo {title} {Elasto-capillary thinning and breakup of model elastic liquids},}\ }\href {\doibase 10.1122/1.1332389} {\bibfield  {journal} {\bibinfo  {journal} {Journal of Rheology}\ }\textbf {\bibinfo {volume} {45}},\ \bibinfo {pages} {115--138} (\bibinfo {year} {2001})}\BibitemShut {NoStop}%
\bibitem [{\citenamefont {McKinley}\ and\ \citenamefont {Sridhar}(2002)}]{McKinley2002}%
  \BibitemOpen
  \bibfield  {author} {\bibinfo {author} {\bibfnamefont {G.~H.}\ \bibnamefont {McKinley}}\ and\ \bibinfo {author} {\bibfnamefont {T.}~\bibnamefont {Sridhar}},\ }\bibfield  {title} {\enquote {\bibinfo {title} {Filament-stretching rheometry of complex fluids},}\ }\href {\doibase 10.1146/annurev.fluid.34.082701.161114} {\bibfield  {journal} {\bibinfo  {journal} {Annual Review of Fluid Mechanics}\ }\textbf {\bibinfo {volume} {34}},\ \bibinfo {pages} {375--415} (\bibinfo {year} {2002})}\BibitemShut {NoStop}%
\bibitem [{\citenamefont {Paredes}, \citenamefont {Tribout},\ and\ \citenamefont {Sepulveda}(1984)}]{paredes1984enthalpies}%
  \BibitemOpen
  \bibfield  {author} {\bibinfo {author} {\bibfnamefont {S.}~\bibnamefont {Paredes}}, \bibinfo {author} {\bibfnamefont {M.}~\bibnamefont {Tribout}}, \ and\ \bibinfo {author} {\bibfnamefont {L.}~\bibnamefont {Sepulveda}},\ }\bibfield  {title} {\enquote {\bibinfo {title} {Enthalpies of micellization of the quaternary tetradecyl-and-cetyl ammonium salts},}\ }\href@noop {} {\bibfield  {journal} {\bibinfo  {journal} {The Journal of Physical Chemistry}\ }\textbf {\bibinfo {volume} {88}},\ \bibinfo {pages} {1871--1875} (\bibinfo {year} {1984})}\BibitemShut {NoStop}%
\bibitem [{\citenamefont {Kadoma}, \citenamefont {Ylitalo},\ and\ \citenamefont {van Egmond}(1997)}]{kadoma1997structural}%
  \BibitemOpen
  \bibfield  {author} {\bibinfo {author} {\bibfnamefont {I.~A.}\ \bibnamefont {Kadoma}}, \bibinfo {author} {\bibfnamefont {C.}~\bibnamefont {Ylitalo}}, \ and\ \bibinfo {author} {\bibfnamefont {J.~W.}\ \bibnamefont {van Egmond}},\ }\bibfield  {title} {\enquote {\bibinfo {title} {Structural transitions in wormlike micelles},}\ }\href@noop {} {\bibfield  {journal} {\bibinfo  {journal} {Rheologica Acta}\ }\textbf {\bibinfo {volume} {36}},\ \bibinfo {pages} {1--12} (\bibinfo {year} {1997})}\BibitemShut {NoStop}%
\bibitem [{\citenamefont {Cooper-White}, \citenamefont {Crooks},\ and\ \citenamefont {Boger}(2002)}]{cooper2002drop}%
  \BibitemOpen
  \bibfield  {author} {\bibinfo {author} {\bibfnamefont {J.}~\bibnamefont {Cooper-White}}, \bibinfo {author} {\bibfnamefont {R.}~\bibnamefont {Crooks}}, \ and\ \bibinfo {author} {\bibfnamefont {D.}~\bibnamefont {Boger}},\ }\bibfield  {title} {\enquote {\bibinfo {title} {A drop impact study of worm-like viscoelastic surfactant solutions},}\ }\href@noop {} {\bibfield  {journal} {\bibinfo  {journal} {Colloids and Surfaces A: Physicochemical and Engineering Aspects}\ }\textbf {\bibinfo {volume} {210}},\ \bibinfo {pages} {105--123} (\bibinfo {year} {2002})}\BibitemShut {NoStop}%
\bibitem [{\citenamefont {Gaonkar}\ and\ \citenamefont {Neuman}(1987)}]{gaonkar1987uncertainty}%
  \BibitemOpen
  \bibfield  {author} {\bibinfo {author} {\bibfnamefont {A.~G.}\ \bibnamefont {Gaonkar}}\ and\ \bibinfo {author} {\bibfnamefont {R.~D.}\ \bibnamefont {Neuman}},\ }\bibfield  {title} {\enquote {\bibinfo {title} {The uncertainty in absolute values of surface tension of water},}\ }\href@noop {} {\bibfield  {journal} {\bibinfo  {journal} {Colloids and Surfaces}\ }\textbf {\bibinfo {volume} {27}},\ \bibinfo {pages} {1--14} (\bibinfo {year} {1987})}\BibitemShut {NoStop}%
\bibitem [{\citenamefont {Hua}\ and\ \citenamefont {Rosen}(1988)}]{hua1988dynamic}%
  \BibitemOpen
  \bibfield  {author} {\bibinfo {author} {\bibfnamefont {X.~Y.}\ \bibnamefont {Hua}}\ and\ \bibinfo {author} {\bibfnamefont {M.~J.}\ \bibnamefont {Rosen}},\ }\bibfield  {title} {\enquote {\bibinfo {title} {Dynamic surface tension of aqueous surfactant solutions: I. basic paremeters},}\ }\href@noop {} {\bibfield  {journal} {\bibinfo  {journal} {Journal of Colloid and Interface Science}\ }\textbf {\bibinfo {volume} {124}},\ \bibinfo {pages} {652--659} (\bibinfo {year} {1988})}\BibitemShut {NoStop}%
\bibitem [{\citenamefont {Rosen}\ and\ \citenamefont {Hua}(1990)}]{rosen1990dynamic}%
  \BibitemOpen
  \bibfield  {author} {\bibinfo {author} {\bibfnamefont {M.~J.}\ \bibnamefont {Rosen}}\ and\ \bibinfo {author} {\bibfnamefont {X.~Y.}\ \bibnamefont {Hua}},\ }\bibfield  {title} {\enquote {\bibinfo {title} {Dynamic surface tension of aqueous surfactant solutions: 2. {P}arameters at 1 s and at mesoequilibrium},}\ }\href@noop {} {\bibfield  {journal} {\bibinfo  {journal} {Journal of Colloid and Interface Science}\ }\textbf {\bibinfo {volume} {139}},\ \bibinfo {pages} {397--407} (\bibinfo {year} {1990})}\BibitemShut {NoStop}%
\bibitem [{\citenamefont {Regev}\ \emph {et~al.}(2010)\citenamefont {Regev}, \citenamefont {Vandebril}, \citenamefont {Zussman},\ and\ \citenamefont {Clasen}}]{regev2010role}%
  \BibitemOpen
  \bibfield  {author} {\bibinfo {author} {\bibfnamefont {O.}~\bibnamefont {Regev}}, \bibinfo {author} {\bibfnamefont {S.}~\bibnamefont {Vandebril}}, \bibinfo {author} {\bibfnamefont {E.}~\bibnamefont {Zussman}}, \ and\ \bibinfo {author} {\bibfnamefont {C.}~\bibnamefont {Clasen}},\ }\bibfield  {title} {\enquote {\bibinfo {title} {The role of interfacial viscoelasticity in the stabilization of an electrospun jet},}\ }\href@noop {} {\bibfield  {journal} {\bibinfo  {journal} {Polymer}\ }\textbf {\bibinfo {volume} {51}},\ \bibinfo {pages} {2611--2620} (\bibinfo {year} {2010})}\BibitemShut {NoStop}%
\bibitem [{\citenamefont {Brust}\ \emph {et~al.}(2013)\citenamefont {Brust}, \citenamefont {Schaefer}, \citenamefont {Doerr}, \citenamefont {Pan}, \citenamefont {Garcia}, \citenamefont {Arratia},\ and\ \citenamefont {Wagner}}]{brust2013rheology}%
  \BibitemOpen
  \bibfield  {author} {\bibinfo {author} {\bibfnamefont {M.}~\bibnamefont {Brust}}, \bibinfo {author} {\bibfnamefont {C.}~\bibnamefont {Schaefer}}, \bibinfo {author} {\bibfnamefont {R.}~\bibnamefont {Doerr}}, \bibinfo {author} {\bibfnamefont {L.}~\bibnamefont {Pan}}, \bibinfo {author} {\bibfnamefont {M.}~\bibnamefont {Garcia}}, \bibinfo {author} {\bibfnamefont {P.}~\bibnamefont {Arratia}}, \ and\ \bibinfo {author} {\bibfnamefont {C.}~\bibnamefont {Wagner}},\ }\bibfield  {title} {\enquote {\bibinfo {title} {Rheology of human blood plasma: Viscoelastic versus newtonian behavior},}\ }\href@noop {} {\bibfield  {journal} {\bibinfo  {journal} {Physical Review Letters}\ }\textbf {\bibinfo {volume} {110}},\ \bibinfo {pages} {078305} (\bibinfo {year} {2013})}\BibitemShut {NoStop}%
\bibitem [{\citenamefont {Rehage}\ and\ \citenamefont {Hoffmann}(1991)}]{rehage1991viscoelastic}%
  \BibitemOpen
  \bibfield  {author} {\bibinfo {author} {\bibfnamefont {H.}~\bibnamefont {Rehage}}\ and\ \bibinfo {author} {\bibfnamefont {H.}~\bibnamefont {Hoffmann}},\ }\bibfield  {title} {\enquote {\bibinfo {title} {Viscoelastic surfactant solutions: model systems for rheological research},}\ }\href@noop {} {\bibfield  {journal} {\bibinfo  {journal} {Molecular Physics}\ }\textbf {\bibinfo {volume} {74}},\ \bibinfo {pages} {933--973} (\bibinfo {year} {1991})}\BibitemShut {NoStop}%
\bibitem [{\citenamefont {Zinelis}\ \emph {et~al.}(2024)\citenamefont {Zinelis}, \citenamefont {Abadie}, \citenamefont {McKinley},\ and\ \citenamefont {Matar}}]{zinelis2024fluid}%
  \BibitemOpen
  \bibfield  {author} {\bibinfo {author} {\bibfnamefont {K.}~\bibnamefont {Zinelis}}, \bibinfo {author} {\bibfnamefont {T.}~\bibnamefont {Abadie}}, \bibinfo {author} {\bibfnamefont {G.~H.}\ \bibnamefont {McKinley}}, \ and\ \bibinfo {author} {\bibfnamefont {O.~K.}\ \bibnamefont {Matar}},\ }\bibfield  {title} {\enquote {\bibinfo {title} {The fluid dynamics of a viscoelastic fluid dripping onto a substrate},}\ }\href@noop {} {\bibfield  {journal} {\bibinfo  {journal} {Soft Matter}\ }\textbf {\bibinfo {volume} {20}},\ \bibinfo {pages} {8198--8214} (\bibinfo {year} {2024})}\BibitemShut {NoStop}%
\bibitem [{\citenamefont {Larson}(1999)}]{Larson1999}%
  \BibitemOpen
  \bibfield  {author} {\bibinfo {author} {\bibfnamefont {R.~G.}\ \bibnamefont {Larson}},\ }\href@noop {} {\emph {\bibinfo {title} {The Structure and Rheology of Complex Fluids}}}\ (\bibinfo  {publisher} {Oxford University Press},\ \bibinfo {year} {1999})\BibitemShut {NoStop}%
\bibitem [{\citenamefont {Okada}, \citenamefont {Urakawa},\ and\ \citenamefont {Inoue}(2016)}]{okada2016reliability}%
  \BibitemOpen
  \bibfield  {author} {\bibinfo {author} {\bibfnamefont {Y.}~\bibnamefont {Okada}}, \bibinfo {author} {\bibfnamefont {O.}~\bibnamefont {Urakawa}}, \ and\ \bibinfo {author} {\bibfnamefont {T.}~\bibnamefont {Inoue}},\ }\bibfield  {title} {\enquote {\bibinfo {title} {Reliability of intrinsic birefringence estimated via the modified stress-optical rule},}\ }\href@noop {} {\bibfield  {journal} {\bibinfo  {journal} {Polymer Journal}\ }\textbf {\bibinfo {volume} {48}},\ \bibinfo {pages} {1073--1078} (\bibinfo {year} {2016})}\BibitemShut {NoStop}%
\bibitem [{\citenamefont {Cerf}\ and\ \citenamefont {Scheraga}(1952)}]{cerf1952flow}%
  \BibitemOpen
  \bibfield  {author} {\bibinfo {author} {\bibfnamefont {R.}~\bibnamefont {Cerf}}\ and\ \bibinfo {author} {\bibfnamefont {H.~A.}\ \bibnamefont {Scheraga}},\ }\bibfield  {title} {\enquote {\bibinfo {title} {Flow birefringence in solutions of macromolecules},}\ }\href@noop {} {\bibfield  {journal} {\bibinfo  {journal} {Chemical Reviews}\ }\textbf {\bibinfo {volume} {51}},\ \bibinfo {pages} {185--261} (\bibinfo {year} {1952})}\BibitemShut {NoStop}%
\bibitem [{\citenamefont {Wunderlich}, \citenamefont {Hoffmann},\ and\ \citenamefont {Rehage}(1987)}]{wunderlich1987flow}%
  \BibitemOpen
  \bibfield  {author} {\bibinfo {author} {\bibfnamefont {I.}~\bibnamefont {Wunderlich}}, \bibinfo {author} {\bibfnamefont {H.}~\bibnamefont {Hoffmann}}, \ and\ \bibinfo {author} {\bibfnamefont {H.}~\bibnamefont {Rehage}},\ }\bibfield  {title} {\enquote {\bibinfo {title} {Flow birefringence and rheological measurements on shear induced micellar structures},}\ }\href@noop {} {\bibfield  {journal} {\bibinfo  {journal} {Rheologica Acta}\ }\textbf {\bibinfo {volume} {26}},\ \bibinfo {pages} {532--542} (\bibinfo {year} {1987})}\BibitemShut {NoStop}%
\bibitem [{\citenamefont {Doi}\ and\ \citenamefont {Edwards}(1986)}]{DoiEdwards1986}%
  \BibitemOpen
  \bibfield  {author} {\bibinfo {author} {\bibfnamefont {M.}~\bibnamefont {Doi}}\ and\ \bibinfo {author} {\bibfnamefont {S.~F.}\ \bibnamefont {Edwards}},\ }\href@noop {} {\emph {\bibinfo {title} {The Theory of Polymer Dynamics}}},\ \bibinfo {series} {International Series of Monographs on Physics}, Vol.~\bibinfo {volume} {73}\ (\bibinfo  {publisher} {Oxford University Press},\ \bibinfo {year} {1986})\BibitemShut {NoStop}%
\bibitem [{\citenamefont {Fuller}(1995)}]{fuller1995optical}%
  \BibitemOpen
  \bibfield  {author} {\bibinfo {author} {\bibfnamefont {G.~G.}\ \bibnamefont {Fuller}},\ }\href@noop {} {\emph {\bibinfo {title} {Optical Rheometry of Complex Fluids}}}\ (\bibinfo  {publisher} {Oxford University Press},\ \bibinfo {year} {1995})\BibitemShut {NoStop}%
\end{thebibliography}%

\end{document}